\documentclass[times,sort&compress,3p]{elsarticle} 
\journal{Journal of Multivariate Analysis}
\usepackage{amsmath, amssymb, enumerate, amsthm}

\usepackage{graphicx,psfrag,epsf}
\usepackage{graphicx,hyperref}
\usepackage{amssymb}
\usepackage{amsmath}
\usepackage{multirow}
\usepackage{float}
\usepackage{mathtools}
\usepackage{xcolor}
\mathtoolsset{showonlyrefs=true}
\usepackage{natbib}
\usepackage{subfig}
\usepackage{lipsum}
\usepackage{graphicx}
\usepackage{epsfig}
\usepackage{epstopdf}
\usepackage{xr}
\usepackage{amssymb}
\journal{ Journal of Multivariate Analysis}

\begin{document}
	
	\begin{frontmatter}
		
		
		
		\title{The joint projected and skew normal; a distribution  for poly-cylindrical data}
		
		
		\author{Gianluca Mastrantonio}
		
		\address{Department of Mathematics, Polytechnic of Turin, Turin, Italy}
		
		\begin{abstract}

			The contribution of this work is the introduction of a multivariate circular-linear (or poly-cylindrical) distribution obtained by combining the projected  and the skew normal.  We show the flexibility of our proposal, its property of closure under marginalization  and how to quantify multivariate dependence.

			Due to a non-identifiability issue that our proposal inherits from the projected normal, a computational problem arises.  
			We overcome it in a Bayesian framework, adding suitable latent variables and showing that  posterior samples can be obtained  with a   post-processing of the  estimation algorithm output.
			Under specific prior choices, this approach enables us to implement a Markov chain Monte Carlo algorithm   relying only on Gibbs steps,   where the updates of the parameters  are done as if we were working with a multivariate normal likelihood. The proposed approach can be also used  with the  projected normal.

			As a proof of concept, on simulated examples we show the ability of our algorithm in recovering the parameters values and to solve the identification problem. Then the proposal is used  in a real data example, where the  turning-angles (circular variables) and the logarithm of the step-lengths (linear variables) of four zebras  are jointly  modelled. 
		\end{abstract}
		
		
	\end{frontmatter}

\section{Introduction}

The analysis of circular data,
i.e., observations with support the unit circle,  requires specific statistical tools since the circular domain  is intrinsically different  from the real line, that is the domain of linear  variables, and this inhibits the use of standard  statistics that if not properly modified   {lead  to} not interpretable results; for a general review  see   \cite{Jammalamadaka2001}, \cite{Merdia1999} or \cite{pewsey2013}. 
A similar type of problem holds for circular densities  that, besides being  non negative and to integrate to 1,  should  possess the property of ``invariance'' \citep{mastrantonio2015f}, i.e.,  {they must be a location model under the group of rotations and reflections of the circle}. This  property,  that expresses the need of densities that can represent equivalently the same phenomena under different reference systems, is peculiar of circular densities and it is  sometimes  overlooked \citep{mastrantonio2015f}.

Circular data are often   observed along with linear ones and 
they are  called  cylindrical if  bivariate, otherwise  poly-cylindrical. For example in marine research wind and wave directions are  modelled with wind speed and wave height \citep{Bulla2012,wang:2014,mastrantonio2016}  and, in  ecology, animal behaviour is described using  measures of    speed and direction, e.g.,  step-length and   turning-angle \citep{DElia2001,Jonsen2005,Patterson2008,Morales2010}. 
In most of the applications  cylindrical data are modelled assuming independence between the circular and linear components, see for example  \cite{Bulla2012}, \cite{lagona2011b} or \cite{morales2004}. Ignoring dependence   can lead to misleading inference     {since we are not considering a component of the data that} can help in understanding the phenomenon under study  \citep[see for example][]{mastrantonio2015}.
In the literature, to date, no  poly-cylindrical  distributions have been proposed  and  there are only
few distributions for cylindrical data; the best known examples are the ones of   \cite{ANDERSONCOOK1997}, \cite{Johnson1978}, \cite{Mardia1978}  and the new density of \cite{Abe2015}.  
The aim of this work is to introduce  what is, to the best of our knowledge, the first poly-cylindrical distribution.

%
%; none of them are suitable  for poly-cylindrical data although  \cite{Johnson1978} provide  a general method  based on copulas  that could be used to construct multivariate circular-linear distributions with specified marginals. 

  {Circular and linear variables live in very different spaces
and the  definition of a mixed-domain distribution is not easy.
The issue is even more complicated if we require flexibility,  interpretable parameters and the possibility to  define  an efficient and  easy to implement  estimation algorithm.  We decide to put ourself in a Bayesian framework because,  as we show in Section \ref{sec:MCMC}, using standard Markov chain Monte Carlo  (MCMC) methods   
we are able to  propose an  algorithm with the required characteristics while Monte Carlo (MC) procedures \citep{brooks2011,Robert2005} allow us to obtain posterior distributions for all the statistics we may need  to describe the results.}

 {Since circular observations show often   bimodality, see for example  \cite{Storch2002} or \cite{Wang2013},  our aim is to propose  a distribution with  circular marginals  that can model such data. In the literature the most known bimodal circular distributions are the projected normal ($\mathcal{PN}$) \citep{Wang2013} and the  generalized von Mises \citep{gatto2007}.  The former  can be easily generalized to the multivariate setting and it has an interesting  augmented density representation, based on a normal  probability density function (pdf), that can be used to define circular-linear dependence. The $\mathcal{PN}$ is  very  flexible   \citep[see for example][]{wang2014,mastrantonio2015b} with shapes that range from  unimodal and symmetric   to bimodal and antipodal,   it is closed under marginalization  and, as we show in the Appendix,  it has the invariance property.    On the other hand, multivariate extensions of the   generalized von Mises are not easy to handle  and, in our opinion, it not straightforward to use it  as a component of a poly-cylindrical distribution. }

 {  We  define our proposal  constructively, starting from the $\mathcal{PN}$ and choosing a distribution for the linear component that, taking advantage of the $\mathcal{PN}$ augmented density representation,   allows us  to  define   a poly-cylindrical distribution whose parameters  can be easily estimated with MCMC algorithms and it is flexible enough to model  real data. }

 { For the linear component we  use a skew normal, that   is a generalization of the Gaussian distribution which allows  more flexibility introducing asymmetry in the normal density. Its first univariate version was proposed by \cite{azzalini85} and following works  introduced  multivariate extensions and different formalizations; see for example \cite{azzalini1996}, \cite{Gupta2004}, \cite{Jones2009} or  \cite{sahu2003}. Among these, we found the one of  \cite{sahu2003}  (hereafter  $\mathcal{SSN}$)  interesting: it can be closed under marginalization and  it has an augmented density representation that, as the $\mathcal{PN}$, is based on a normal pdf. }

 %and linear components that are, respectively,   the  projected normal \citep{Wang2013}  and  the skew normal of \cite{sahu2003} (hereafter $\mathcal{PN}$ and $\mathcal{SSN}$), and  introducing dependence between the two components. 

%The $\mathcal{SSN}$ is a generalization of the Gaussian distribution which allows  more flexibility introducing asymmetry in the normal density. Its first univariate version was proposed by \cite{azzalini85} and following works  introduced  multivariate extensions and different formalizations, see for example \cite{azzalini1996}, \cite{Gupta2004} or  \cite{sahu2003};  among these we  use the one of \cite{sahu2003}  that, under specific set of parameters,  is closed under marginalization. 

 {Using this particular form of the skew normal distribution, due to the properties listed above,  we are able to  define  the \emph{joint projected and skew normal} ($\mathcal{JPSN}$) poly-cylindrical distribution by introducing  dependence in the normal pdfs of the augmented representations.   The distribution retains  the $\mathcal{PN}$ and $\mathcal{SSN}$ as  marginal distributions and is
 closed under marginalization, i.e., any subset of circular and linear variables is   $\mathcal{JPSN}$ distributed.  The MCMC algorithm we propose can be based only on   Gibbs steps, updating   parameters  as if we were working with  a multivariate normal likelihood. The density cannot be  expressed in closed form but, from the point of view of model fitting, since we are able to estimate its parameters easily we do not consider this an issue.  }

  { The $\mathcal{JPSN}$ has the same identification problem of the $\mathcal{PN}$ \citep{Wang2013}, but we show that  posterior values  can be obtained  by a    post-processing of the MCMC algorithm based on  the  non-identifiable likelihood. The proposed  algorithm can be also used   with the univariate and multivariate  $\mathcal{PN}$   and  the spherical $\mathcal{PN}$ distribution
 	of \cite{Stumpfhause2016}, solving their identification problem in a new way.    }

The algorithm, tested on simulated datasets,  shows  its ability in retrieving the parameters used to simulate the data and posterior samples do not suffer from an  identification issue. We used the $\mathcal{JPSN}$ to  jointly model  the logarithm of step-lengths  and turning-angles of 4 zebras observed in Botswana (Africa).
 {A comparison  based on the continuous rank probability scores (CRPSs) \cite{Gneiting2007,grimit2006}  between our proposal, the cylindrical distribution of \cite{Abe2015}  and a cylindrical version of the  $\mathcal{JPSN}$, i.e.,  assuming independence between zebras,  is  provided, showing that  ignoring multivariate dependence can lead to  loss of predictive ability.}

The paper is organized as follows. Section \ref{sec:joints} is devoted to the constructive definition  of  the distribution. In Section \ref{sec:idsolve}   we introduce the identification problem and how to estimate the $\mathcal{JPSN}$ parameters. The proposal  is applied to  simulated examples in Section \ref{sec:sim} and the  real data application is shown in  Section \ref{sec:real2}.
The paper ends with  concluding remarks in Section \ref{sec:conc}.  In the Appendix  we prove the invariance property of the $\mathcal{PN}$ and  we show  MCMC implementation details.

\section{The  joint projected and skew normal distribution} \label{sec:joints}

In this section  we build the poly-cylindrical density  by  
 first introducing the circular and linear marginals and then showing how to induce dependence.

\subsection{The   projected normal distribution} \label{sec:multi} \label{sec:multi2}

\newsavebox{\MuUno}
\savebox{\MuUno}{$\boldsymbol{\mu}_{w_i}=\left(\begin{smallmatrix}2\\0\end{smallmatrix}\right)$}
\newsavebox{\SigmaUno}
\savebox{\SigmaUno}{$\boldsymbol{\Sigma}_{w_i}=\left(\begin{smallmatrix}1&0\\0&1\end{smallmatrix}\right)$}

\newsavebox{\MuDue}
\savebox{\MuDue}{$\boldsymbol{\mu}_{w_i}=\left(\begin{smallmatrix}2\\0\end{smallmatrix}\right)$}
\newsavebox{\SigmaDue}
\savebox{\SigmaDue}{$\boldsymbol{\Sigma}_{w_i}=\left(\begin{smallmatrix}1&0.9\\0.9&1\end{smallmatrix}\right)$}

\newsavebox{\MuTre}
\savebox{\MuTre}{$\boldsymbol{\mu}_{w_i}=\left(\begin{smallmatrix}-0.1\\-0.2\end{smallmatrix}\right)$}
\newsavebox{\SigmaTre}
\savebox{\SigmaTre}{$\boldsymbol{\Sigma}_{w_i}=\left(\begin{smallmatrix}1&-0.9\\-0.9&1\end{smallmatrix}\right)$}

%\savebox{\smlmat}{$\left(\begin{smallmatrix}2&-2\\-1&2\end{smallmatrix}\right)$}

\begin{figure}[t!]
	\centering 
	{\subfloat[]{\includegraphics[trim= {0.7cm 1.cm 0.7cm 1.cm},clip,scale=0.24]{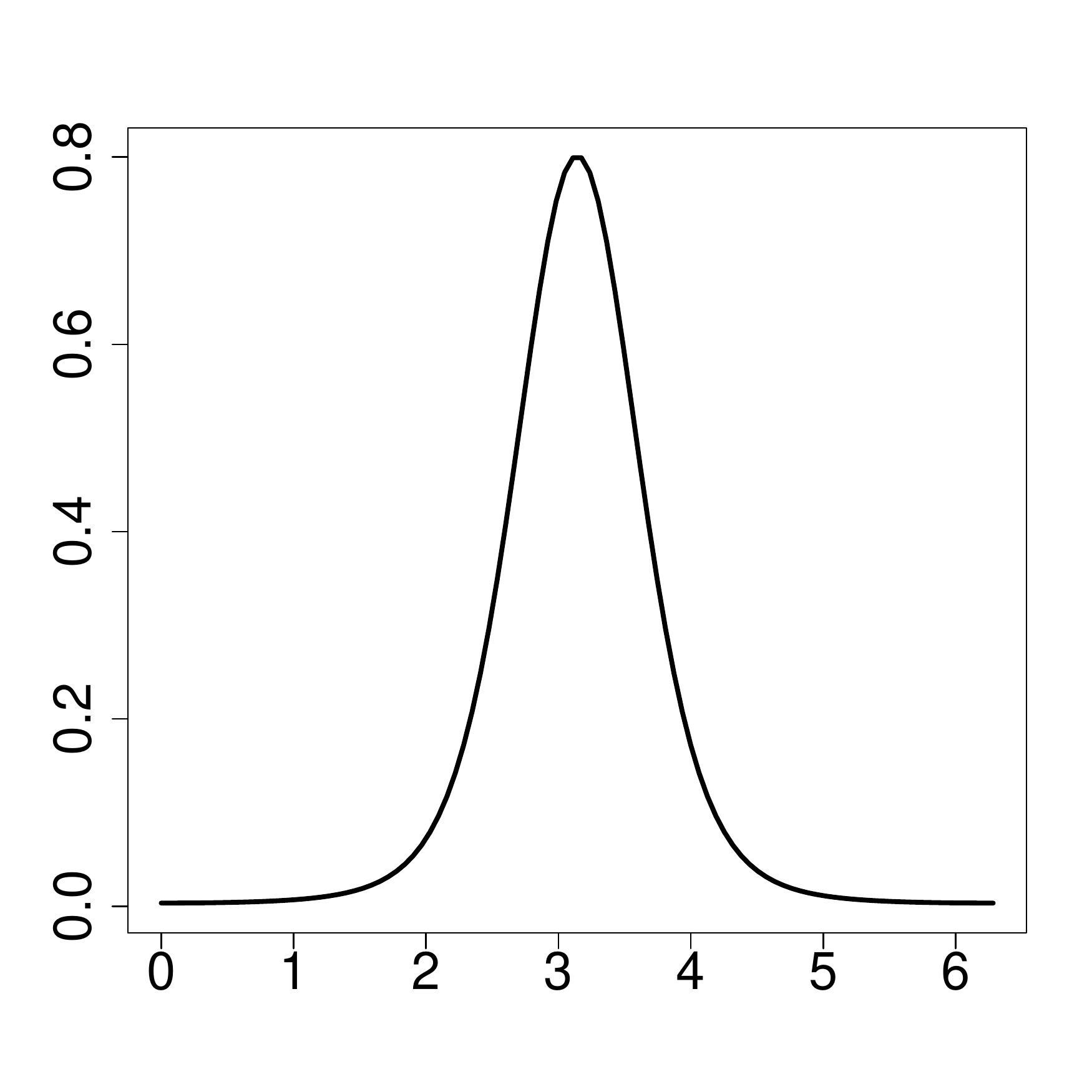}}}
	{\subfloat[]{\includegraphics[trim= {0.7cm 1.cm 0.7cm 1.cm},clip,scale=0.24]{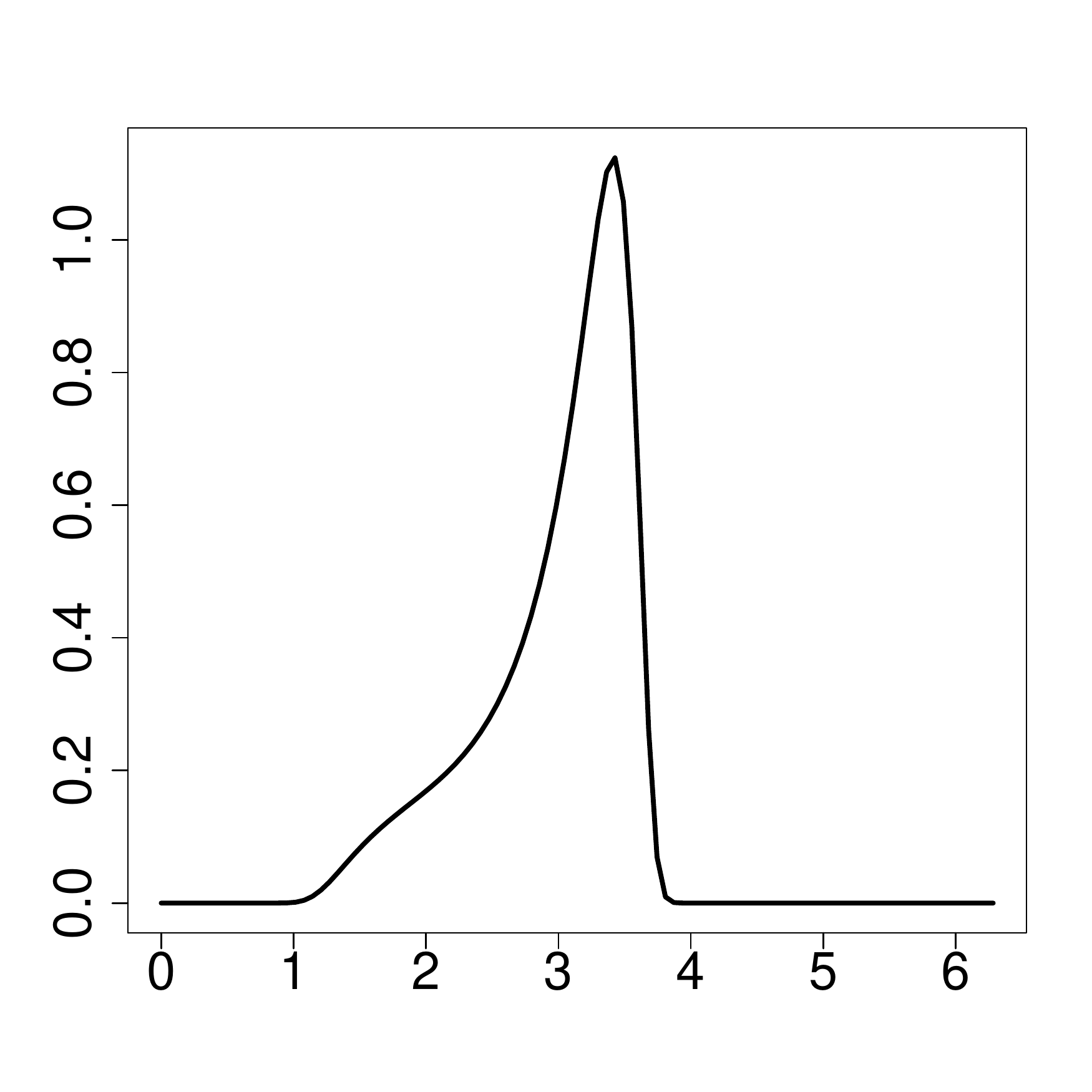}}}
	{\subfloat[]{\includegraphics[trim= {0.7cm 1.cm 0.7cm 1.cm},clip,scale=0.24]{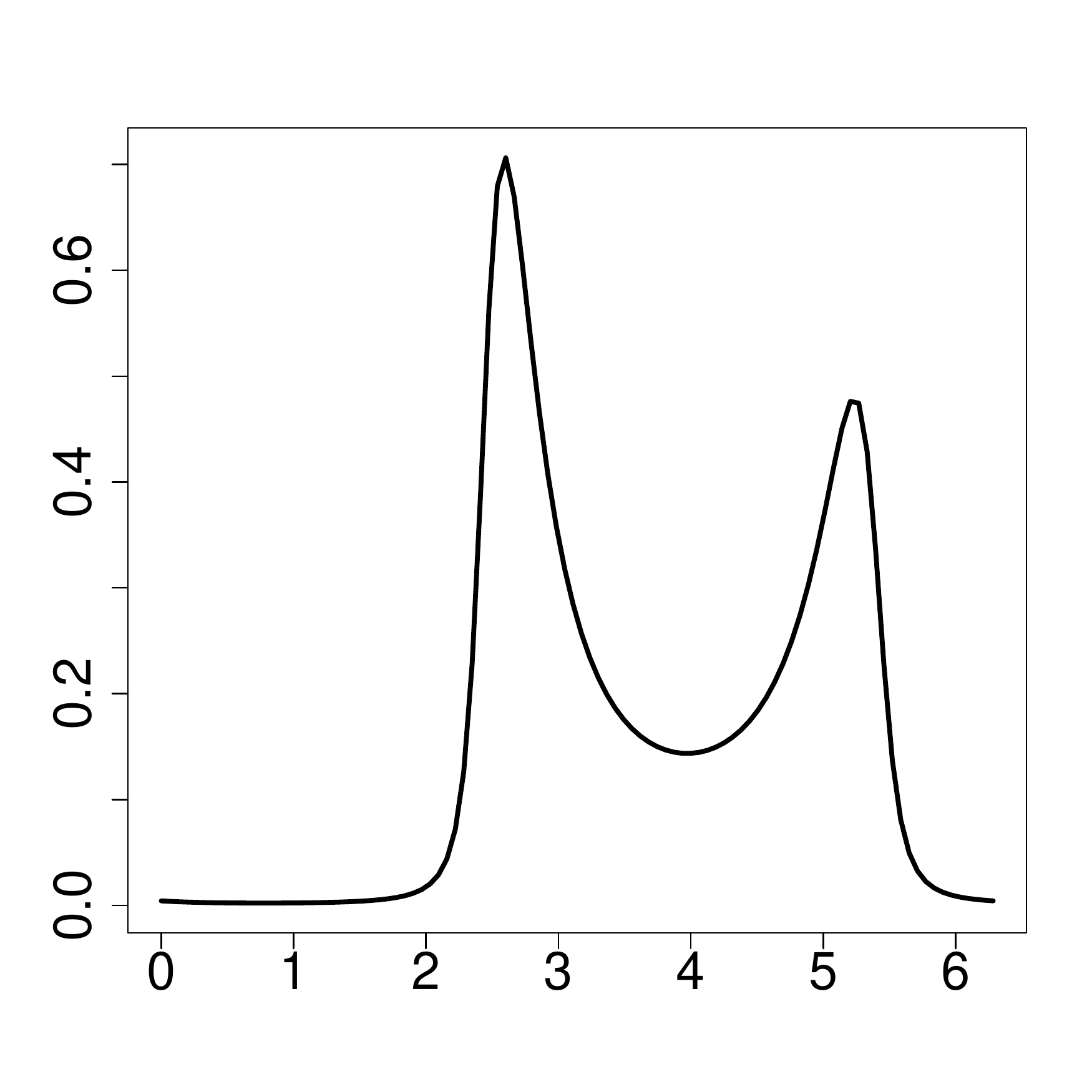}}}
	\caption{Univariate projected normal densities under three sets of  parameters:  (a) \usebox{\MuUno}, \usebox{\SigmaUno}; (b) \usebox{\MuDue}, \usebox{\SigmaDue}; (c) \usebox{\MuTre}, \usebox{\SigmaTre} }  \label{fig:denssss}
\end{figure}

 {The $\mathcal{PN}$  is a distribution for a $p$-dimensional vector  $\boldsymbol{\Theta}=\{\Theta_i\}_{i=1}^p$ of circular variables, i.e., $\Theta_i \in [0,2\pi)$ is an angle expressed in radiant,}  obtained starting from  a $2p$-dimensional vector   $\mathbf{W}= \{ \mathbf{W}_{i} \}_{i=1}^{p}$, where $\mathbf{W}_i = (W_{i1},W_{i2})^{\top}\in \mathbb{R}^2$, distributed as a $2p$-variate normal with mean vector   $\boldsymbol{\mu}_w$ and covariance matrix $\boldsymbol{\Sigma}_w$.
$\mathbf{W}_i$, normally distributed with parameters $\{\boldsymbol{\mu}_{w_i},\boldsymbol{\Sigma}_{w_i}\}$, is a point  in the 2-dimensional space expressed using the Cartesian system.  The same point can be also  represented in  polar coordinates   with  the angle $\Theta_i$ and the distance vector $R_i\in \mathbb{R}^+$. Between $\mathbf{W}_i$,   $\Theta_i$ and $R_i$ the following relations exist:
\begin{equation} \label{eq:tranpn}
\Theta_i = \text{ atan}^* \left(\frac{W_{i2}}{W_{i1}}\right)
\end{equation} 
 {and
\begin{equation} 	\label{eq:u}
\mathbf{W}_i=R_{i} \left(
\begin{array}{c}
\cos \Theta_i\\
\sin \Theta_i
\end{array}
\right) ,\, R_i= ||\mathbf{W}_i ||, 
\end{equation}
where 
\begin{equation}
\text{ atan}^* \left(\frac{S}{C}\right) =
\begin{cases}
\text{ atan}\left(\frac{S}{C}  \right) & \mbox{if } C>0, \mbox{ }S \geq 0,\\
\frac{\pi}{2} & \mbox{if } C=0, \mbox{ }S > 0,\\
\text{ atan}\left(\frac{S}{C}  \right)  + \pi & \mbox{if } C<0, \\
\text{ atan}\left(\frac{S}{C}  \right)+2 \pi & \mbox{if } C\geq 0, \mbox{ }S < 0,\\
\mbox{undefined} & \mbox{if } C=0, \mbox{ }S = 0,\\
\end{cases}
\end{equation}
 is a modified  arctangent function used to define  a quadrant-specific inverse of the tangent.}

If we transform each $\mathbf{W}_i$ in $(\Theta_i,R_i)^{\top}$ the Jacobian of the transformation is $\prod_{i=1}^p R_i$ and then the joint density  of $(\boldsymbol{\Theta},\mathbf{R})^{\top}$, where  $\mathbf{R}= \{R_i\}_{i=1}^p$,  is  given by
\begin{equation} \label{eq:multipn}
f(\boldsymbol{\theta},\mathbf{r})= \prod_{i=1}^p r_{i} \phi_{2p}(\mathbf{w}|\boldsymbol{\mu}_w,\boldsymbol{\Sigma}_w),
\end{equation}
where $f(\cdot)$ indicates the density of its arguments, $\mathbf{r}$ is a realization of $\mathbf{R}$,  $\phi_{2p}(\mathbf{w}|\boldsymbol{\mu}_w,\boldsymbol{\Sigma}_w)$ is the pdf  evaluated at $\mathbf{w}$ of a $2p-$variate normal distribution with mean  $\boldsymbol{\mu}_w$ and covariance matrix $\boldsymbol{\Sigma}_w$; here $\mathbf{w}$ must be seen as a function of $(\boldsymbol{\theta},\mathbf{r})^{\top}$.

The marginal density of  $\boldsymbol{\Theta}$, obtained by  integrating out $\mathbf{R}$ in  \eqref{eq:multipn}, is a  $p-$variate projected normal    with parameters $\boldsymbol{\mu}_w$ and $\boldsymbol{\Sigma}_w$, i.e., $\boldsymbol{\Theta} \sim \mathcal{PN}_{p}(\boldsymbol{\mu}_w, \boldsymbol{\Sigma}_w)$.  As shown in \cite{Wang2013}, the $\mathcal{PN}$ can be symmetric,  asymmetric and 
bimodal; univariate   shapes are depicted in Figure \ref{fig:denssss}.

 {A closed form expression for the $\mathcal{PN}_p$ density is only available in the univariate case ($p=1$) and it is
\begin{align}
f(\theta_i)  &= \frac{\phi_2( \boldsymbol{\mu}_{w_i}|\mathbf{0}_2, \boldsymbol{\Sigma}_{w_i})+|\boldsymbol{\Sigma}_{w_i}|^{-1} D(\theta_i)      \Phi_1(D(\theta_i) |0,1 )   \phi_1 \left(    |\boldsymbol{\Sigma}_{w_i}|^{-1}  C(\theta_i)^{-1/2}  \left(  \mu_{w_{i1}} \sin \theta_i-   \mu_{w_{i2}} \cos \theta_i       \right)     \right)     }{  C(\theta_i) } , \\
%a &= |\boldsymbol{\Sigma}_{w_i}|^{-1},\\
\end{align}
where
\begin{align}
C(\theta_i) & = |\boldsymbol{\Sigma}_{w_i}|^{-2} \left( \sigma_{w_{i2}}^2 \cos^2 \theta_i-  \rho_{w_i} \sigma_{w_{i1}}\sigma_{w_{i2}} \sin 2\theta_i +    \sigma_{w_{i1}}^2 \sin^2 \theta_i    \right),\\
D(\theta_i) & =  \frac{   |\boldsymbol{\Sigma}_{w_i}|^{-2}  \left(   \mu_{w_{i1}} \sigma_{w_{i2}} \left( \sigma_{w_{i2}} \cos \theta_i-   \rho_{w_i}     \sigma_{w_{i1}} \sin \theta_i \right)   +       \mu_{w_{i2}} \sigma_{w_{i1}} \left( \sigma_{w_{i1}} \sin \theta_i-   \rho_{w_i}     \sigma_{w_{i2}} \cos \theta_i \right)                 \right)   }{\sqrt{C(\theta_i)}},
\end{align}
$\Phi_{\ell}(\cdot|\cdot,\cdot)$ indicates the normal $\ell$-variate  cumulative distribution function with given mean vector and covariance matrix, $\mathbf{0}_{\ell}$ is a vector of 0s of length $\ell$,  $\mu_{w_{ij}}$ and $\sigma_{w_{ij}}^2$ are the mean and variance of $W_{ij}$ and $\rho_{w_i} $ is the correlation between $W_{i1}$ and $W_{i2}$.}

%\citep[see for example][]{Wang2013} and

In practical applications \citep[see for example][]{wang2014,mastrantonio2015b,Maruotti2015a} it is  generally preferable to work with the pair  $(\boldsymbol{\Theta},\mathbf{R})^{\top}$ that has the nice closed form density given  in equation \eqref{eq:multipn}, treating $\mathbf{R}$ as a vector of latent variables.

%
%Following \cite{Merdia1999}, we say that $\Theta_i$ has projected normal distribution with parameters  $(\boldsymbol{\mu}_{w_i},\boldsymbol{\Sigma}_{w_i})^{\top}$, i.e. $\Theta_i \sim PN(\boldsymbol{\mu}_{w_i},\boldsymbol{\Sigma}_{w_i})$. The marginal density of $\Theta_i$, obtained  integrating out $R_i$ in  \eqref{eq:teta},  has an almost intractable closed form.  In practical applications \citep[see for example][]{wang2014,mastrantonio2015b,Maruotti2015a} it is  generally preferable to work with the couple  $(\Theta_i,R_i)^{\top}$ that has the nice closed form density in equation \eqref{eq:teta}, treating $R_i$ as latent variable.   Some of the  PN shapes are depicted in Figure \ref{fig:denssss}. 

 { The multivariate $\mathcal{PN}$ is closed under marginalization since  $\boldsymbol{\Theta}_{A} \sim \mathcal{PN}_{n_a}(\boldsymbol{\mu}_{w,A},\boldsymbol{\Sigma}_{w,A})$, where $A \subset \{1,\dots , p\}$, $n_a$ indicates the cardinality of the set $A$ and  $\{\boldsymbol{\mu}_{w,A},\boldsymbol{\Sigma}_{w,A}\}$ are  mean and covariance matrix of $\mathbf{W}_{A}$.   Moreover, as we show in Appendix  \ref{sec:app1}, the univariate $\mathcal{PN}$ posses the invariance property and then inference does not depend on the reference system chosen for the circular variables   \citep{mastrantonio2015f}.}

\subsection{The skew normal distribution} \label{sec:skew}

\begin{figure}[t!]
	\centering 
	{\subfloat[]{\includegraphics[trim= {0.7cm 1.cm 0.7cm 1.cm},clip,scale=0.24]{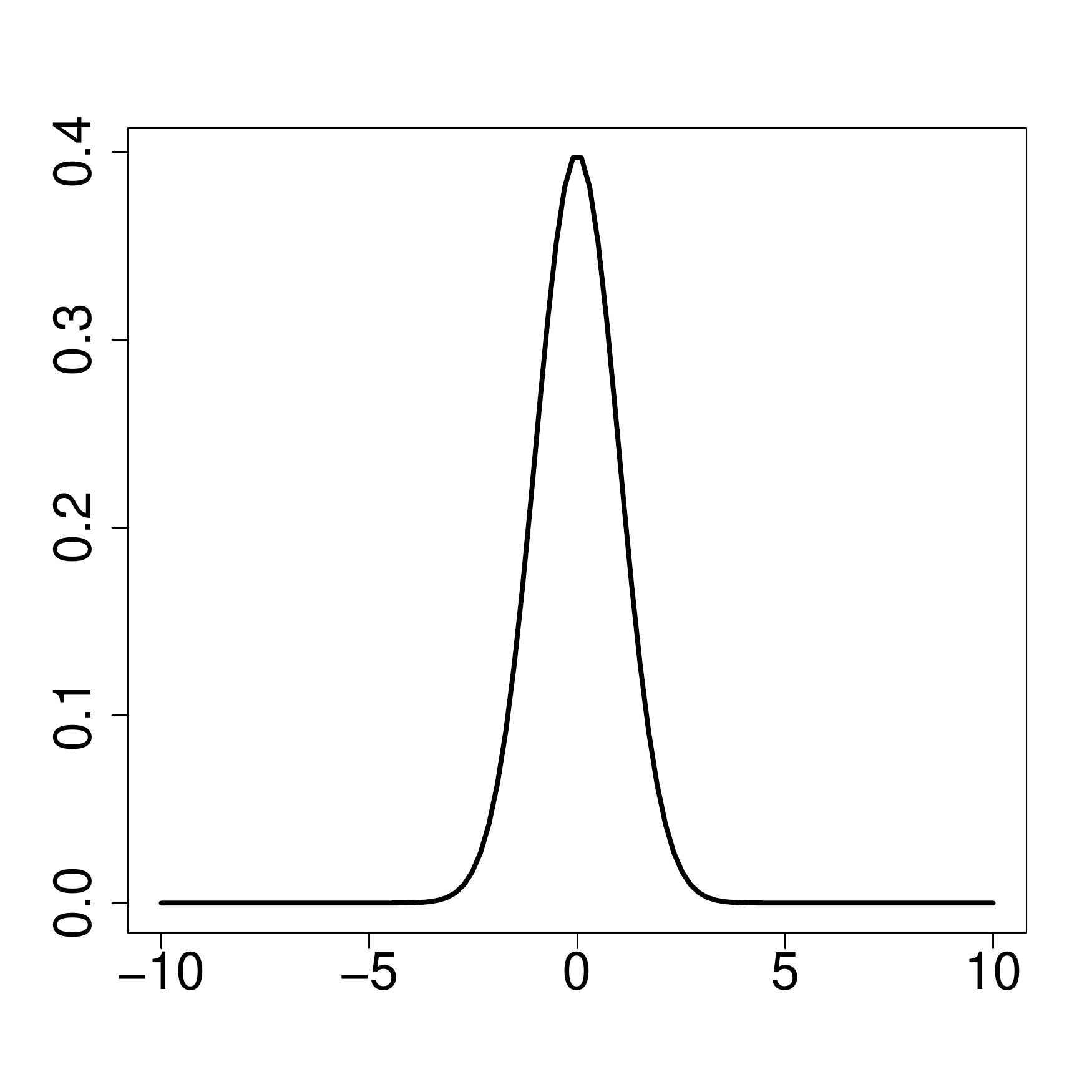}}}
	{\subfloat[]{\includegraphics[trim= {0.7cm 1.cm 0.7cm 1.cm},clip,scale=0.24]{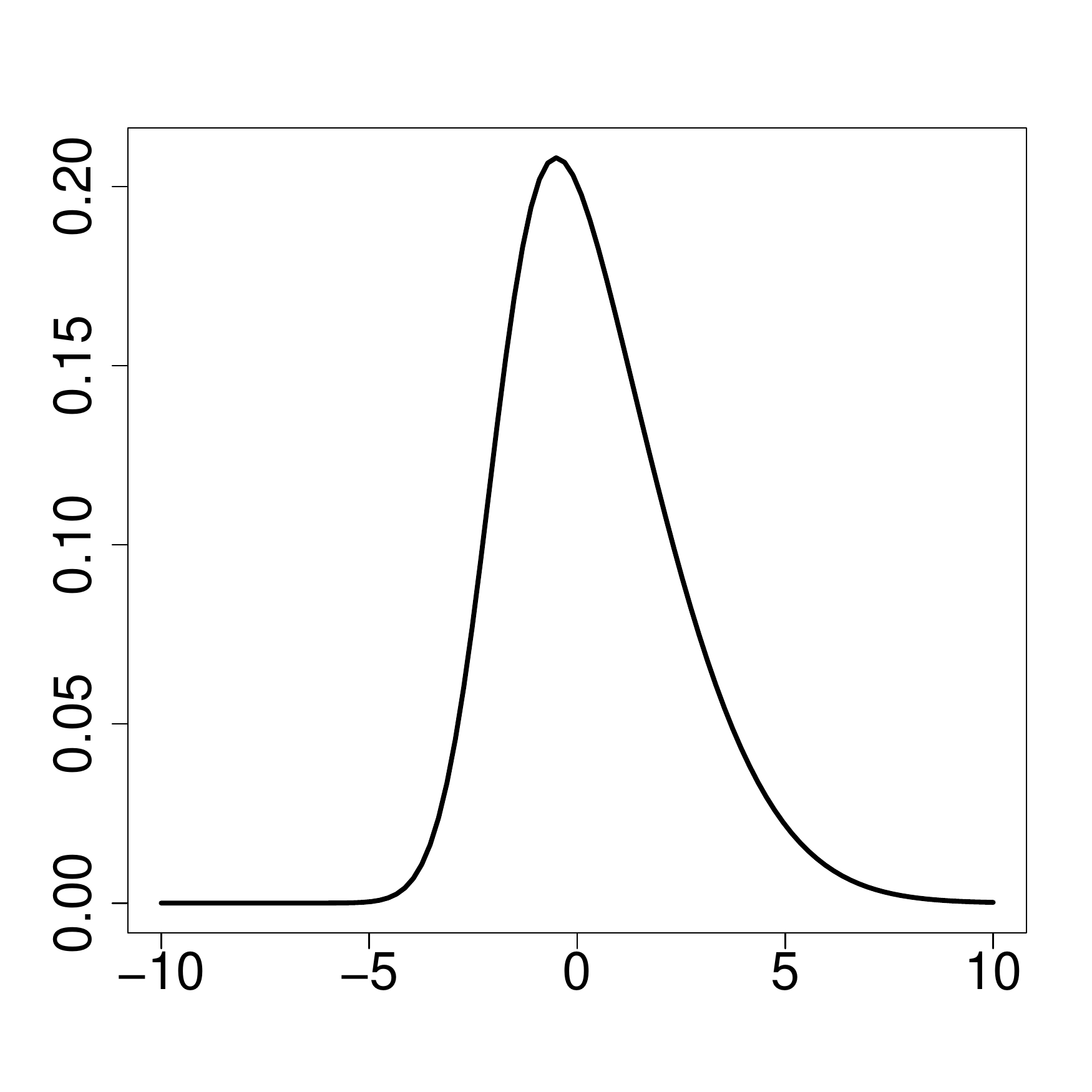}}}
	{\subfloat[]{\includegraphics[trim= {0.7cm 1.cm 0.7cm 1.cm},clip,scale=0.24]{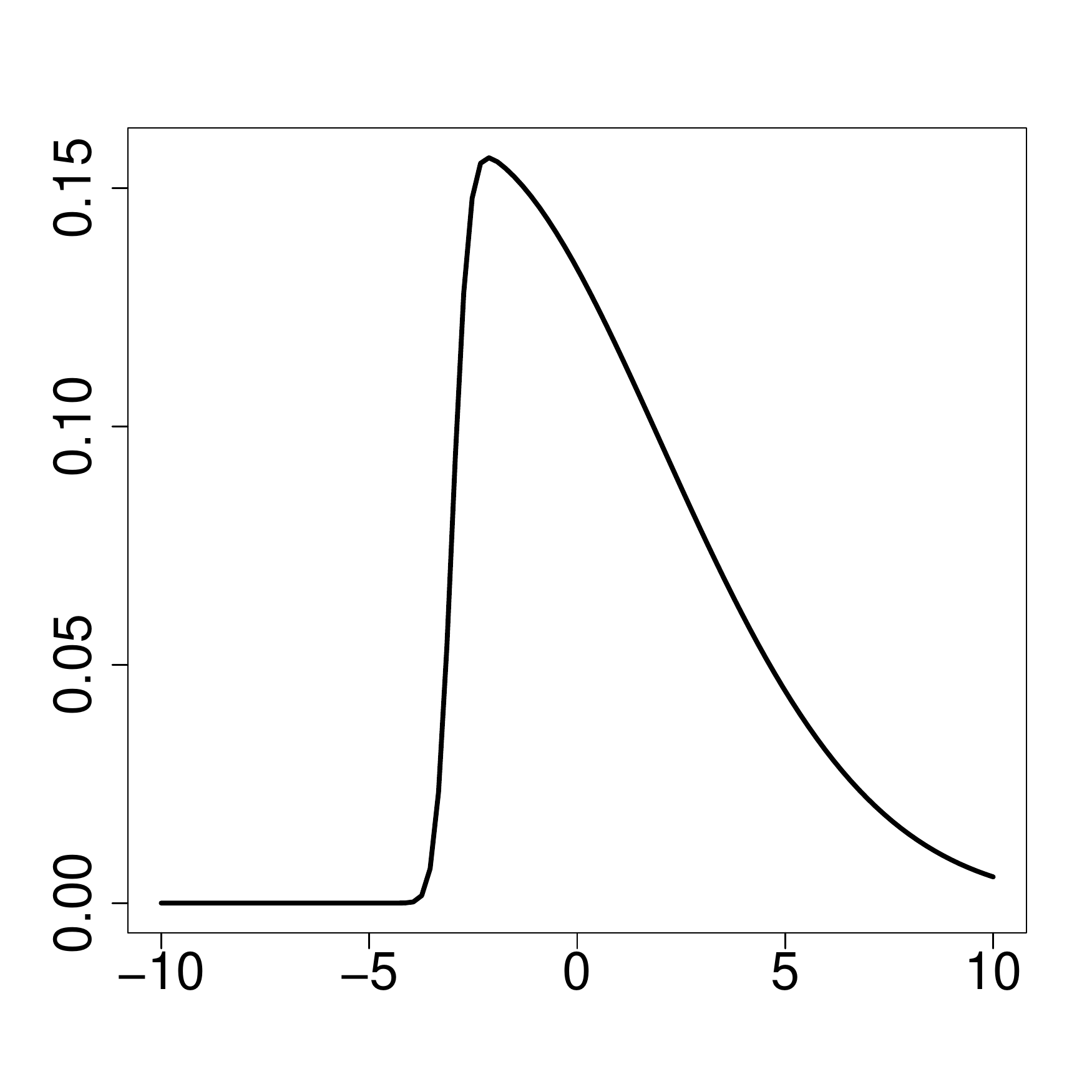}}}
	\caption{Univariate skew normal densities under three sets of parameters: 
	(a) $\boldsymbol{\mu}_y=0$, $\boldsymbol{\Sigma}_{y}=1$, $\boldsymbol{\lambda}=0$; 
	(b) $\boldsymbol{\mu}_y=-2$, $\boldsymbol{\Sigma}_{y}=1$, $\boldsymbol{\lambda}=3$; 
	(c) $\boldsymbol{\mu}_y=-3$, $\boldsymbol{\Sigma}_{y}=0.1$, $\boldsymbol{\lambda}=5$.}  \label{fig:densssss}
\end{figure}

We now introduce the skew normal distribution of \cite{sahu2003} as the distribution of a $q-$dimensional vector $\boldsymbol{Y} = \{Y_{j} \}_{j=1}^q$, with $Y_j \in \mathbb{R}$. 
Let $\boldsymbol{\mu}_{y}$ be a vector of length $q$, $\boldsymbol{\Sigma}_y$ be a $q \times q$ non-negative definite (nnd) matrix and  {$\boldsymbol{\Lambda} = \mbox{diag}(\boldsymbol{\lambda})$ be a $q \times q$ diagonal matrix with diagonal elements   $\boldsymbol{\lambda}=\{\lambda_{i}  \}_{i=1}^q \in \mathbb{R}^q$}.   We say that $\boldsymbol{Y}$ is distributed accordingly to a  $q-$variate skew normal  with parameters $\boldsymbol{\mu}_{y}$,  $\boldsymbol{\Sigma}_y$ and $\boldsymbol{\lambda}$ ($\boldsymbol{Y} \sim \mathcal{SSN}_{q}(\boldsymbol{\mu}_{y}, \boldsymbol{\Sigma}_y,\boldsymbol{\lambda})$) if it has pdf 
\begin{equation} \label{eq:skew}
f(\mathbf{y}) = 	2^q\phi_q \left(\mathbf{y}| \boldsymbol{\mu}_{y}, \boldsymbol{\Upsilon} \right) \Phi_q \left(\boldsymbol{\Lambda}^{\top}\boldsymbol{\Upsilon}^{-1}(\mathbf{y}-\boldsymbol{\mu}_y)|\mathbf{0}_q, \boldsymbol{\varGamma}  \right),
\end{equation}
where   $\boldsymbol{\Upsilon} = \boldsymbol{\Sigma}_{y}+\boldsymbol{\Lambda}\boldsymbol{\Lambda}^{\top}$,   $\boldsymbol{\varGamma} = \mathbf{I}_q-\boldsymbol{\Lambda}^{\top}\boldsymbol{\Upsilon}^{-1}\boldsymbol{\Lambda}$ and $\mathbf{I}_q$ is the identity matrix of dimension $q$.
% and 
% we assume a diagonal $\boldsymbol{\Lambda}$, i.e.  $\boldsymbol{\Lambda} = \mbox{diag}(\boldsymbol{\lambda})$ with   $\boldsymbol{\lambda}=\{\lambda_{i}  \}_{i=1}^q$, since  in this case the $\mathcal{SSN}$ is closed under marginalization (see equation \eqref{eq:skewrep}) and then the same property will hold for our poly-cylindrical distribution (see next section). 
 {Although in  \cite{sahu2003} $\boldsymbol{\Lambda}$ is defined as a full matrix, here we constrain it to be diagonal to have a  $\mathcal{SSN}$  closed under marginalization; the same property will be inherited by our poly-cylindrical distribution (see Section \ref{sec:joint}). 
From 	 \eqref{eq:skew} we clearly see that $\mathbf{Y} \sim \mathcal{N}_{q}(\boldsymbol{\mu}_{y}, \boldsymbol{\Sigma}_y)$ if $\boldsymbol{\Lambda}$   is a null matrix and for this reason  it  is  called the  \emph{skew parameter}. } 
Examples of univariate $\mathcal{SSN}$ densities are shown in Figure \ref{fig:densssss}.

The $\mathcal{SSN}$ has a nice stochastic representation \citep{arellano2007} that is useful for the definition of the poly-cylindrical distribution.
Let  $\mathbf{D} \sim \mathcal{HN}_q(\mathbf{0}_q, \mathbf{I}_q)$, where $ \mathcal{HN}_q(\cdot,\cdot)$ indicates the $q-$dimensional half normal \citep{Olmos2012},
%, i.e. a truncated normal defined over $\mathbb{R}^+$, 
and  $\mathbf{H} \sim \mathcal{N}_q(\mathbf{0}_q, \boldsymbol{\Sigma}_{y})$, then  $\mathbf{Y}$ can be written as 
\begin{equation} \label{eq:skewrep}
	\mathbf{Y} = \boldsymbol{\mu}_{y} + \boldsymbol{\Lambda} \mathbf{D}+\mathbf{H}.
\end{equation}

From \eqref{eq:skewrep} we can see that  $\mathbf{Y}|\mathbf{D}= \mathbf{d}$ is normally distributed with mean $ \boldsymbol{\mu}_{y} + \boldsymbol{\Lambda} \mathbf{d}$ and covariance matrix  $\boldsymbol{\Sigma}_{y}$. Consequently, the joint density of $(\mathbf{Y}, \mathbf{D})^{\top}$ expressed as the product of the ones of $\mathbf{Y}|\mathbf{D}$ and $\mathbf{D}$, is  given by
\begin{equation} \label{eq:laty}
f(\mathbf{y},\mathbf{d})=2^q\phi_{q}( \mathbf{y}  |\boldsymbol{\mu}_y+\boldsymbol{\Lambda}\mathbf{d} , {\boldsymbol{\Sigma}}_y) \phi_q(\mathbf{d}| \mathbf{0}_q,\mathbf{I}_q ). 
\end{equation}

%
%
%
%We want our circular-linear distribution to be closed under marginalization. As will be clear in the next section, this is  only possible  if the  SSN has the same property and this  can be achieved by assuming, as we are going to do from now on,   $\boldsymbol{\Lambda} = \mbox{diag}(\boldsymbol{\lambda})$ with   $\boldsymbol{\lambda}=\{\lambda_{i}  \}_{i=1}^q$.  

\subsection{The joint linear-circular distribution} \label{sec:joint}

\begin{figure}[t!]
	\centering 
	\captionsetup[subfigure]{labelformat=empty}
	    \subfloat{\raisebox{+0.68in}{\rotatebox[origin=t]{90}{Example 1}}}
	{\subfloat{\includegraphics[trim= {0.7cm 1.cm 0.7cm 1.cm},clip,scale=0.22]{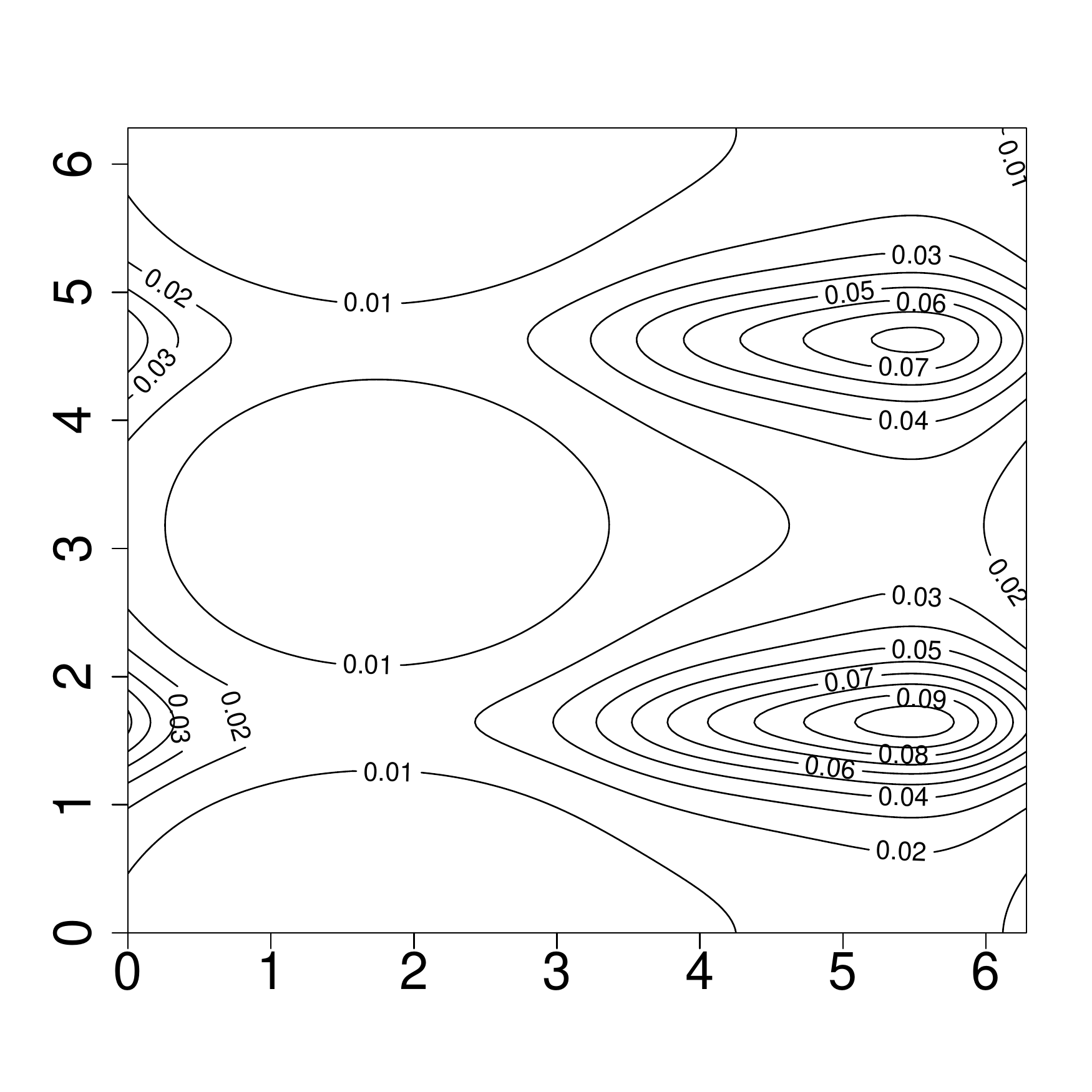}}}
	{\subfloat{\includegraphics[trim= {0.7cm 1.cm 0.7cm 1.cm},clip,scale=0.22]{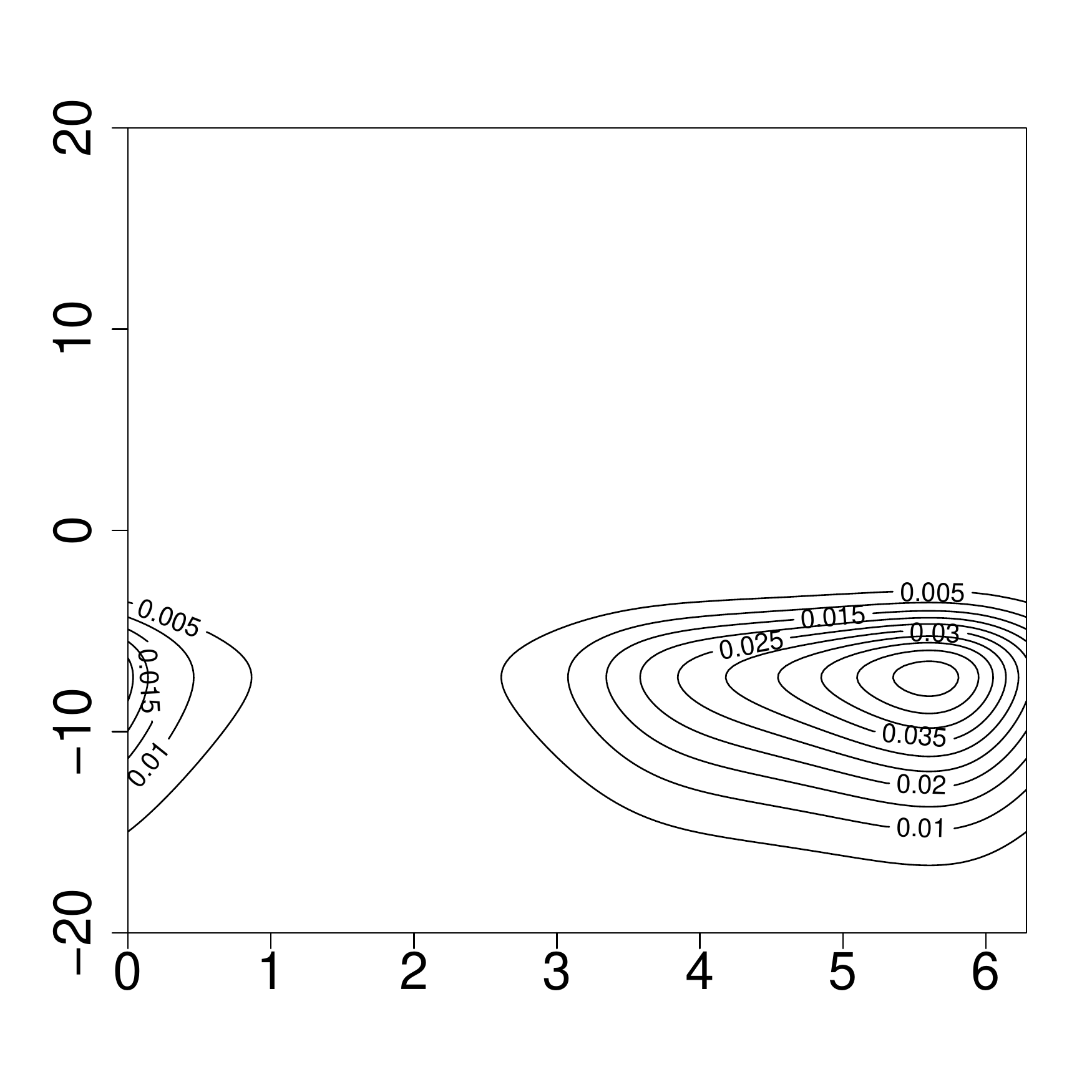}}}
	{\subfloat{\includegraphics[trim= {0.7cm 1.cm 0.7cm 1.cm},clip,scale=0.22]{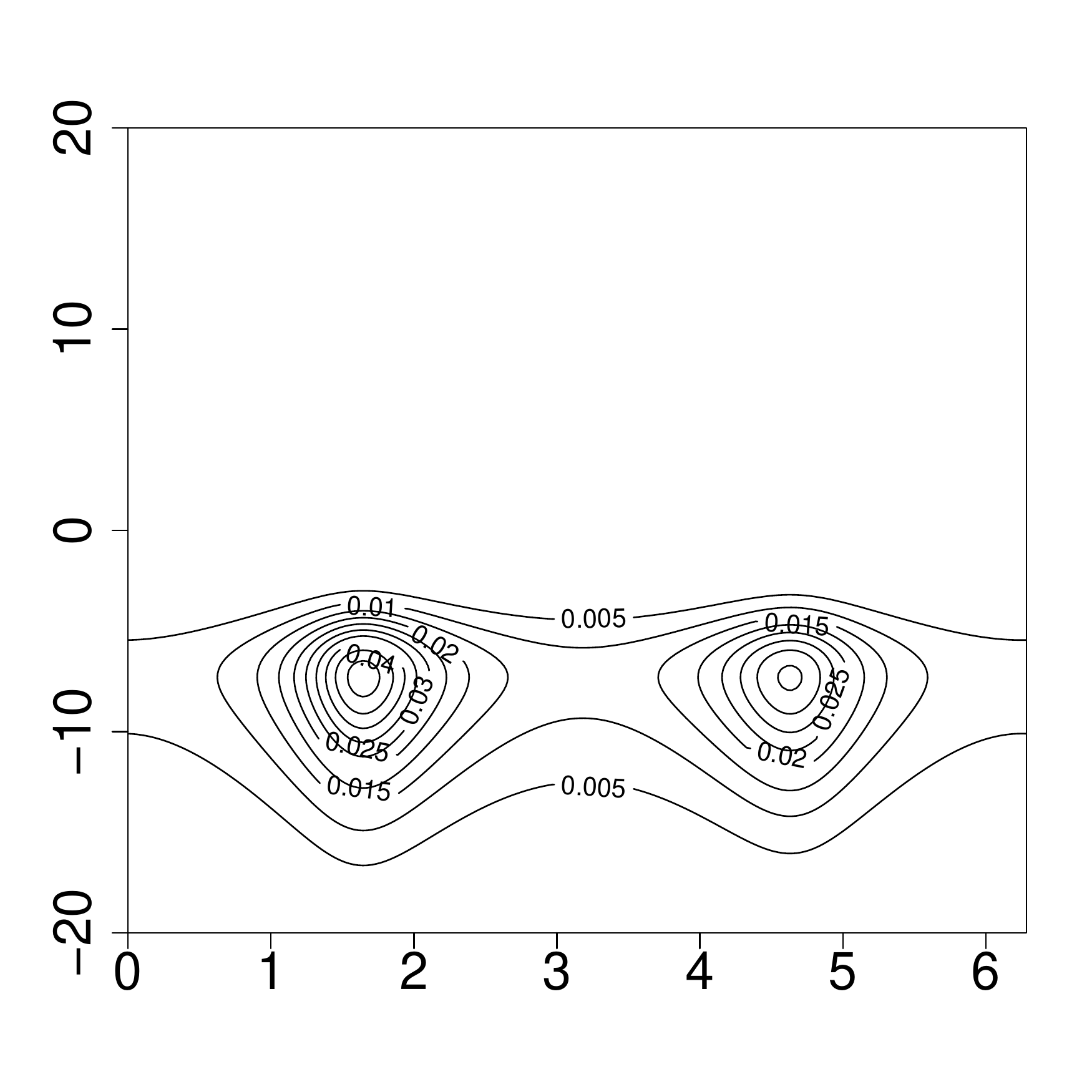}}}\\
	\subfloat{\raisebox{+0.68in}{\rotatebox[origin=t]{90}{Example 2}}}
	{\subfloat{\includegraphics[trim= {0.7cm 1.cm 0.7cm 1.cm},clip,scale=0.22]{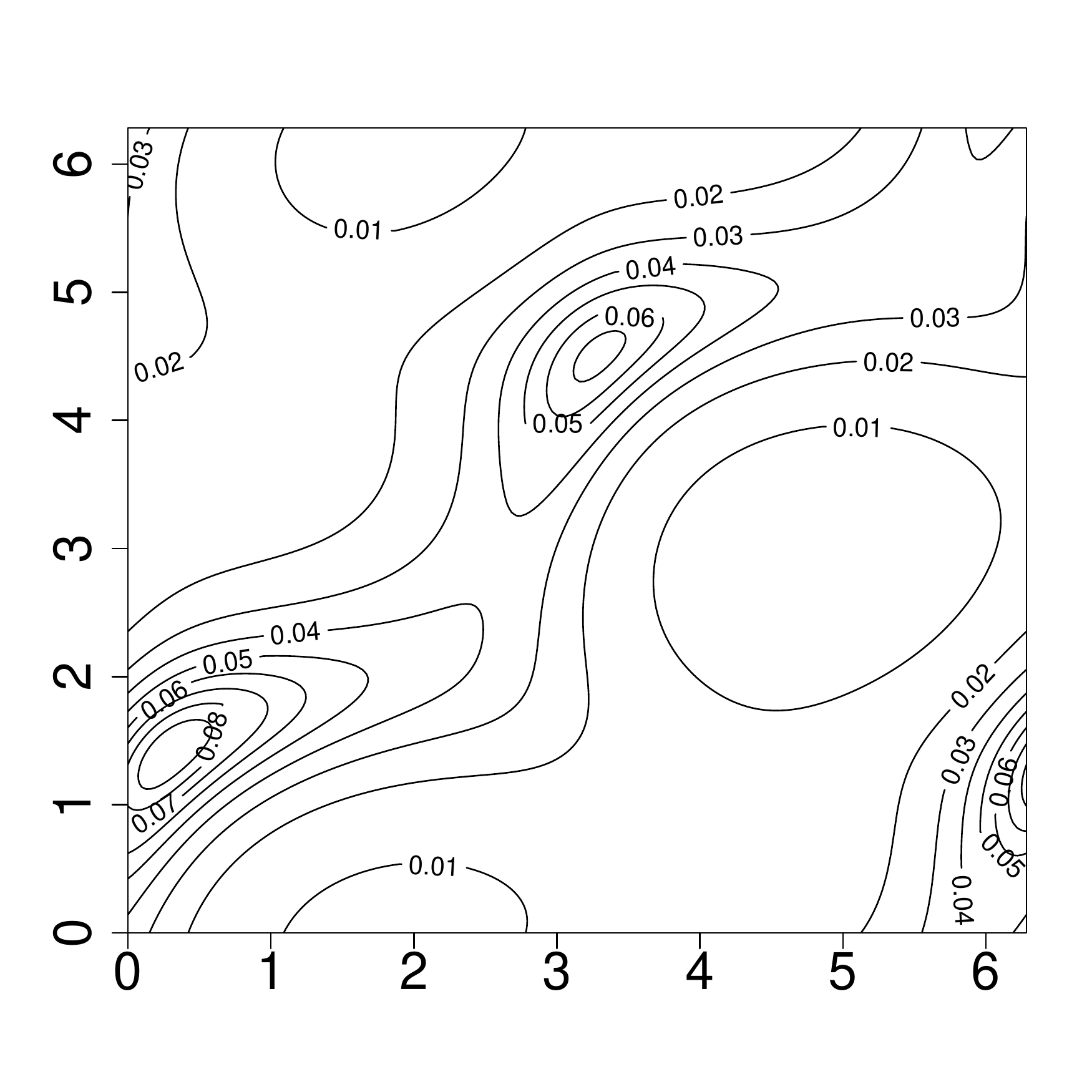}}}
	{\subfloat{\includegraphics[trim= {0.7cm 1.cm 0.7cm 1.cm},clip,scale=0.22]{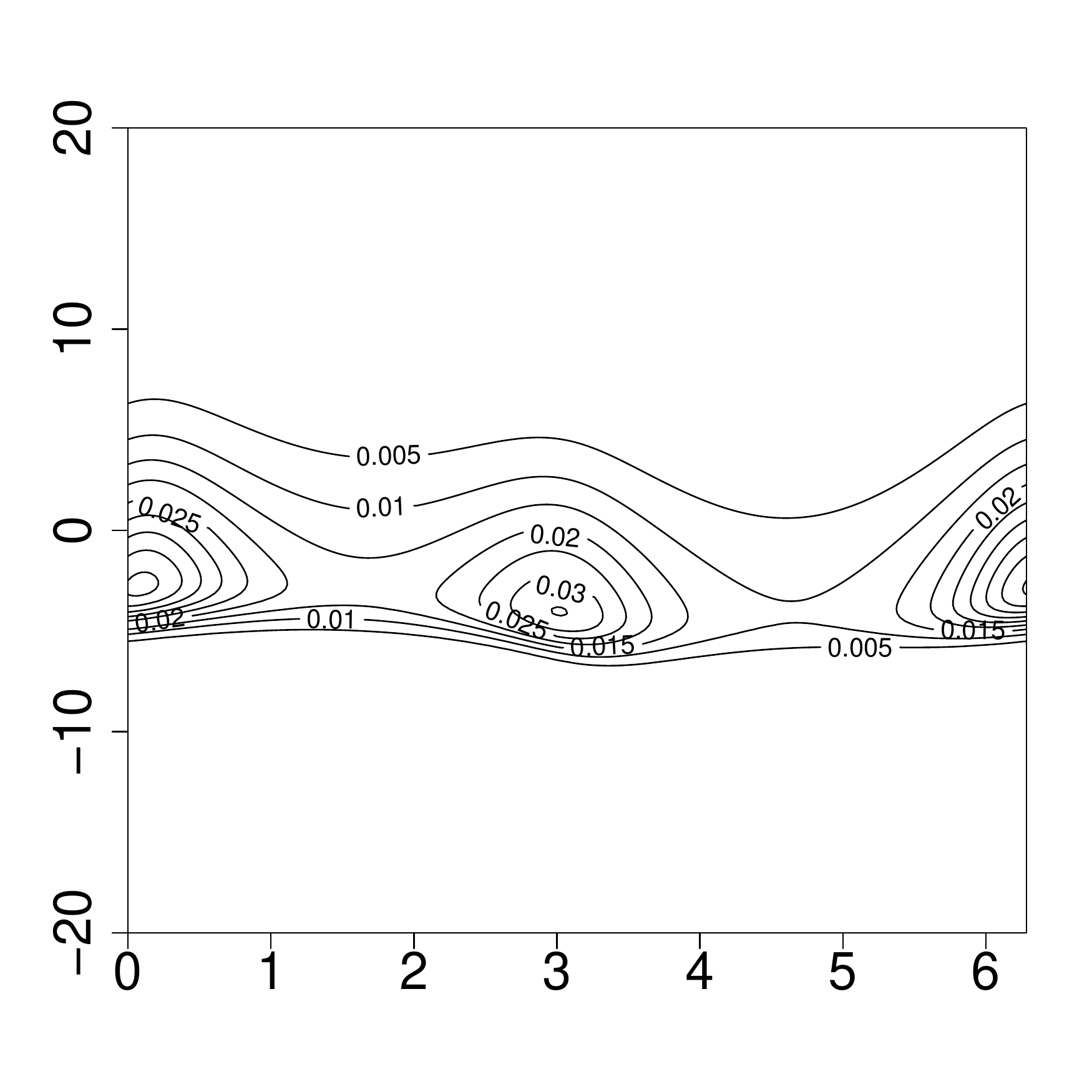}}}
	{\subfloat{\includegraphics[trim= {0.7cm 1.cm 0.7cm 1.cm},clip,scale=0.22]{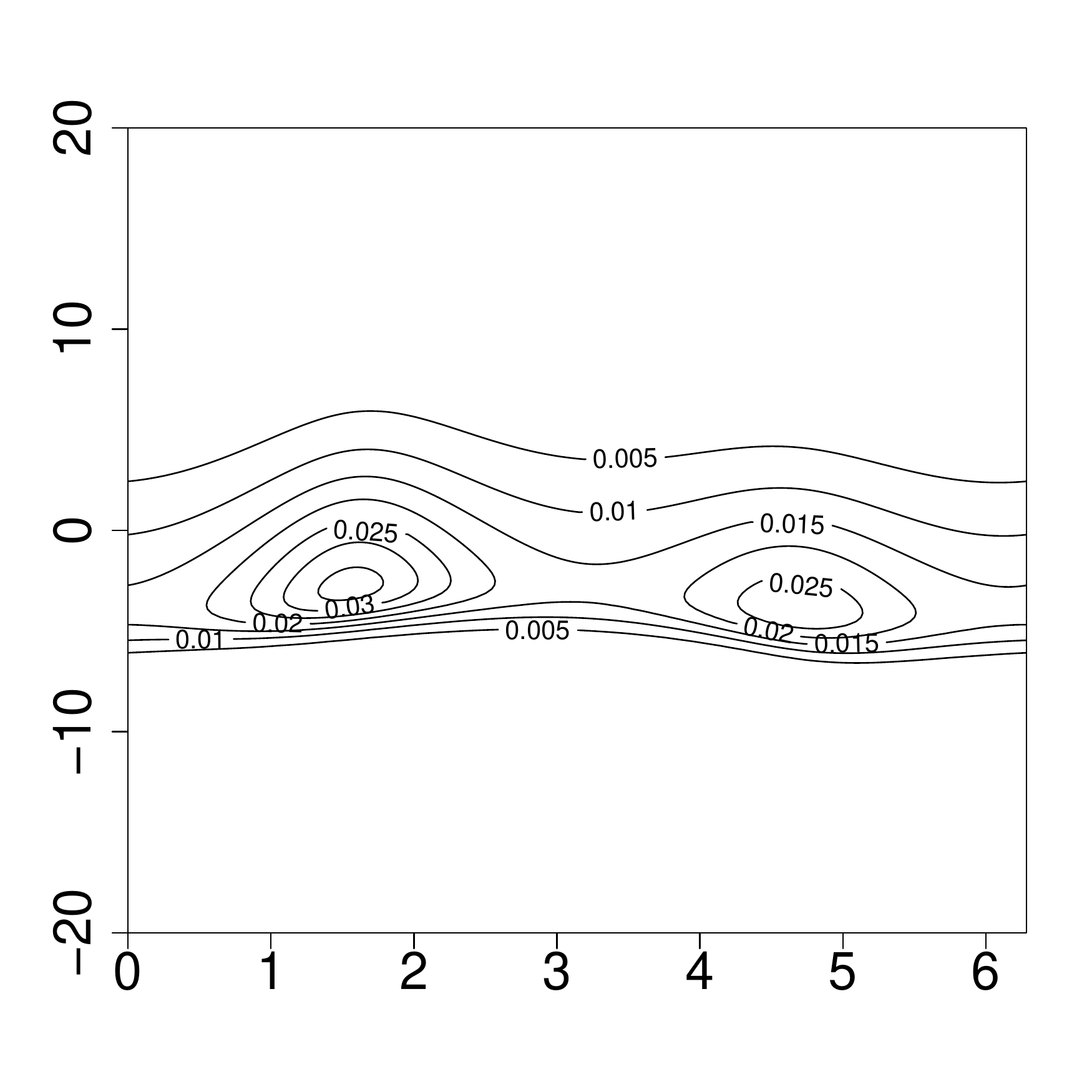}}}\\
	\subfloat{\raisebox{+0.68in}{\rotatebox[origin=t]{90}{Example 3}}}
	{\subfloat[$(\Theta_1,\Theta_2)$]{\includegraphics[trim= {0.7cm 1.cm 0.7cm 1.cm},clip,scale=0.22]{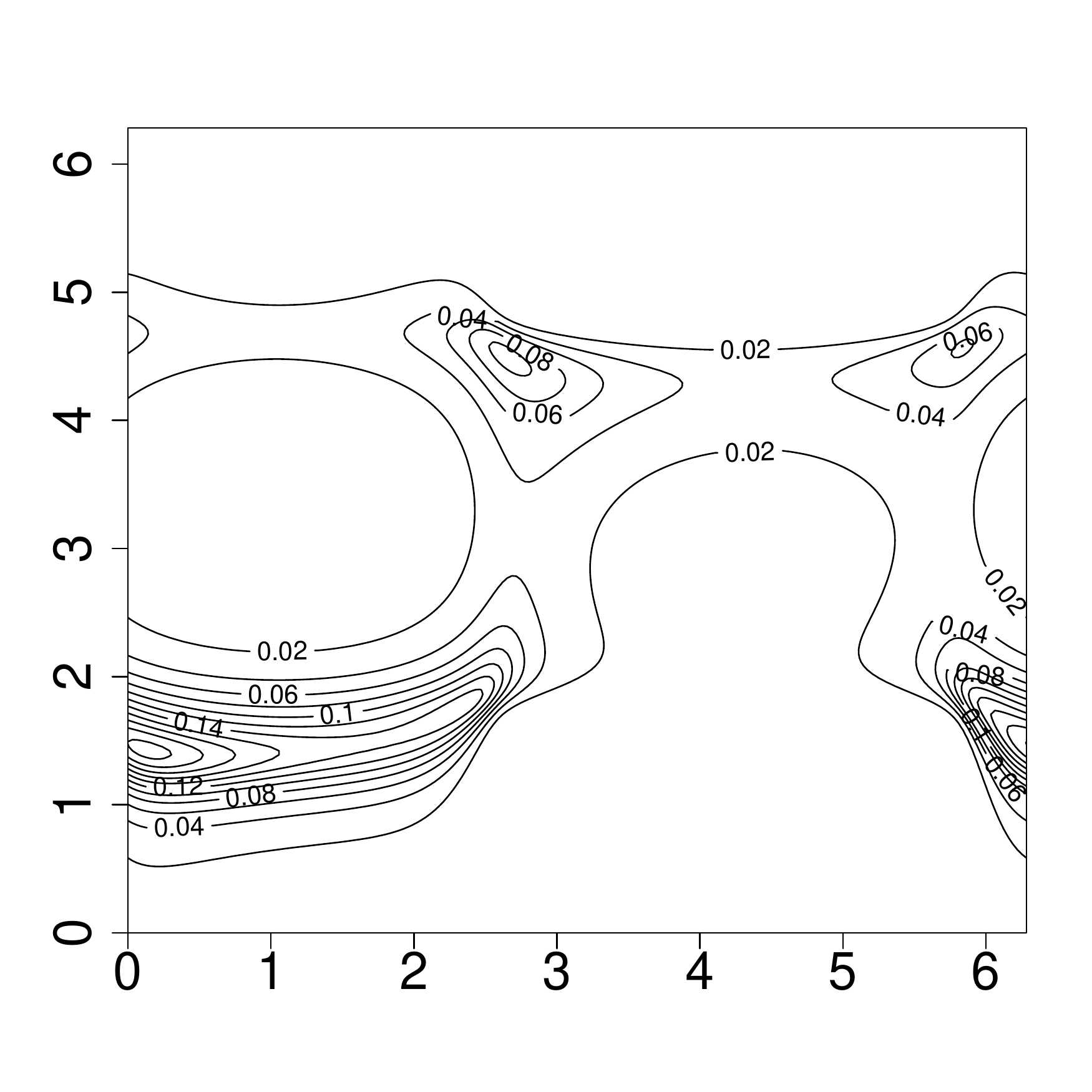}}}
	{\subfloat[$(\Theta_1,Y_1)$]{\includegraphics[trim= {0.7cm 1.cm 0.7cm 1.cm},clip,scale=0.22]{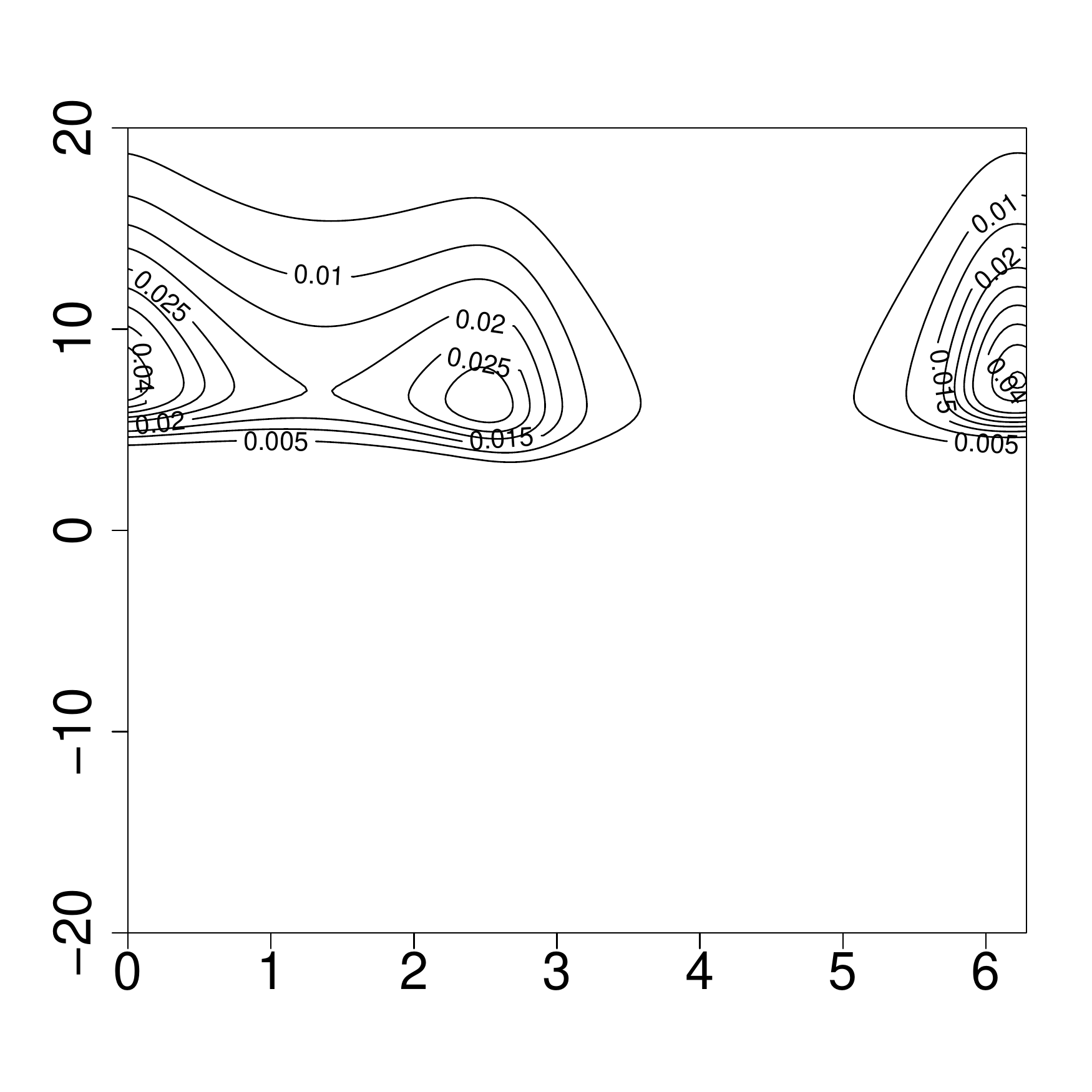}}}
	{\subfloat[$(\Theta_2,Y_1)$]{\includegraphics[trim= {0.7cm 1.cm 0.7cm 1.cm},clip,scale=0.22]{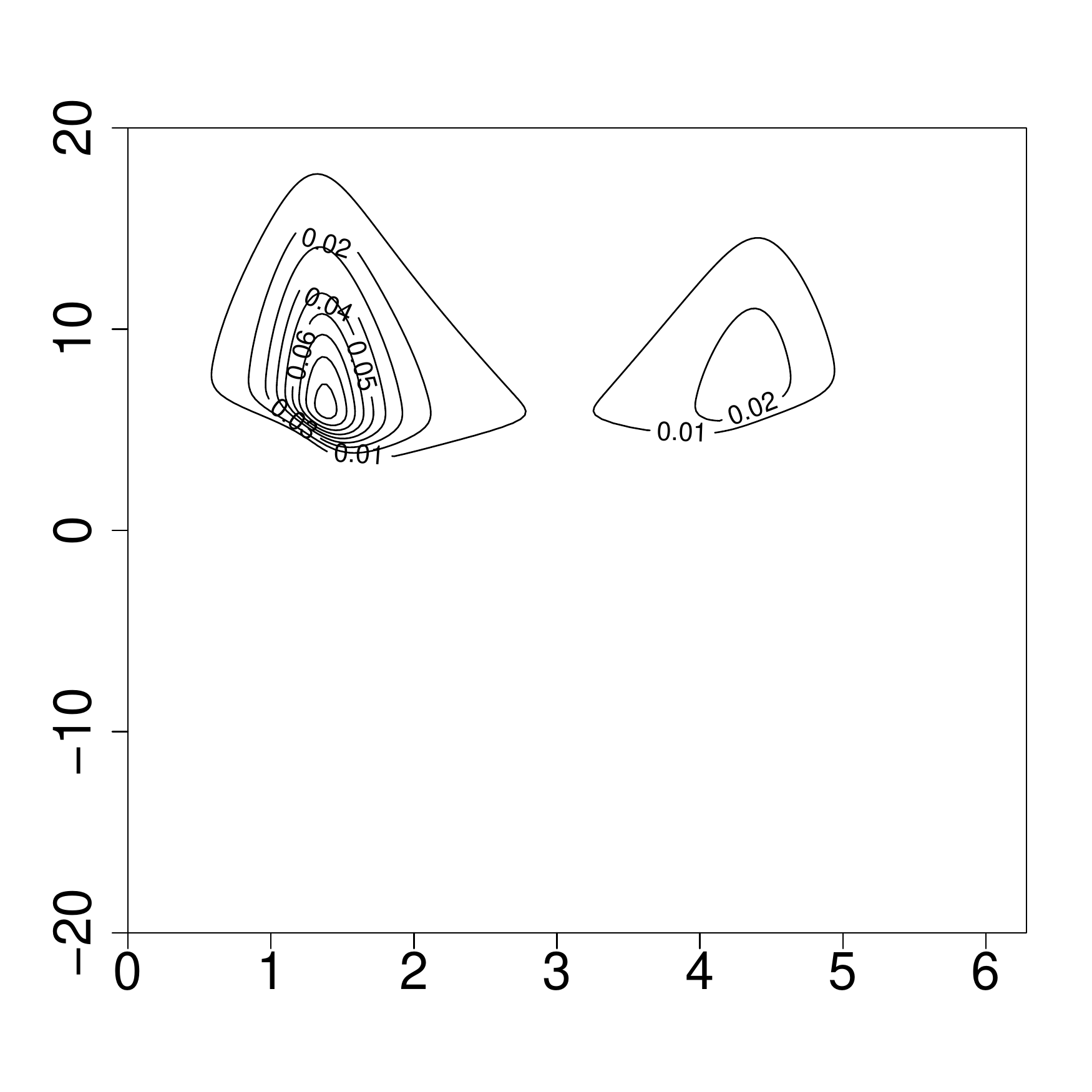}}}
	\caption{Bivariate marginal distributions  of a $\mathcal{JPSN}_{2,1}(\boldsymbol{\mu},\boldsymbol{\Sigma},\boldsymbol{\lambda})$ under three sets of parameters (row) reported in Section \ref{sec:sim}. In the first column there are the marginal distributions of $(\Theta_1,\Theta_2)$, in the second those of $(\Theta_1,Y_1)$ and in the third  the marginals of $(\Theta_2,Y_1)$.}  \label{fig:dens}
\end{figure}

 {
In this section we   define the poly-cylindrical distribution starting from the augmented circular and linear marginals shown in equations \eqref{eq:multipn} and \eqref{eq:laty}. As in the  previous sections, we indicate with $p$  and $q$ the dimensions of the vectors of circular and linear variables, respectively.}

% If we assume independence between  circular and linear components,   the joint density of $(\boldsymbol{\Theta},\mathbf{R},\mathbf{Y},\mathbf{D})^{\top}$ is the product of \eqref{eq:multipn} and \eqref{eq:laty}:
%\begin{equation}
%f(\boldsymbol{\theta},\mathbf{r},\mathbf{y},\mathbf{d})=2^q \phi_{2p}(\mathbf{w}|\boldsymbol{\mu}_w,\boldsymbol{\Sigma}_w)\phi_{q}( \mathbf{y}  |\boldsymbol{\mu}_y+\text{diag}(\boldsymbol{\lambda})\mathbf{d} , {\boldsymbol{\Sigma}}_y) \phi_q(\mathbf{d}| \mathbf{0}_q,\mathbf{I}_q ) \prod_{i=1}^p r_{i}.
%\end{equation}
%Marginalization over $(\mathbf{R},\mathbf{D})^{\top}$ gives  the density of $(\boldsymbol{\Theta},\mathbf{Y})^{\top}$,  that is not available  in closed form as the  $\mathcal{PN}_p$  density.

 {
It is natural to introduce dependence between $\boldsymbol{\Theta}$ and $\mathbf{Y}$ by substituting the two normal pdfs of the augmented representation, i.e.,  $\phi_{2p}(\mathbf{w}|\boldsymbol{\mu}_w,\boldsymbol{\Sigma}_w)$  and $ \phi_{q}( \mathbf{y}  |\boldsymbol{\mu}_y+\text{diag}(\boldsymbol{\lambda})\mathbf{d} , {\boldsymbol{\Sigma}}_y)$, with a $2p+q$ normal pdf that has the two pdfs as marginals;  after marginalization we obtain the density of $(\boldsymbol{\Theta},\mathbf{Y})^{\top}$.} More precisely, we define the joint density of  $(\boldsymbol{\Theta},\mathbf{R},\mathbf{Y},\mathbf{D})^{\top}$ as 
\begin{equation} \label{eq:ssq}
 f(\boldsymbol{\theta},{\mathbf{r}},\mathbf{y},\mathbf{d})= 2^q\phi_{2p+q}((\mathbf{w}, \mathbf{y} )^{\top} |\boldsymbol{\mu}+(\mathbf{0}_{2p},\text{diag}(\boldsymbol{\lambda})\mathbf{d} )^{\top} , {\boldsymbol{\Sigma}}) \phi_q(\mathbf{d}| \mathbf{0}_q,\mathbf{I}_q ) \prod_{i=1}^p r_{i},
\end{equation}
with 
$\boldsymbol{\mu} = (\boldsymbol{\mu}_w,\boldsymbol{\mu}_y)^{\top}$ and 
$$ %\begin{equation}
\boldsymbol{\Sigma}=\left(
\begin{array}{cc}
\boldsymbol{\Sigma}_{w} & \boldsymbol{\Sigma}_{wy}\\
\boldsymbol{\Sigma}_{wy}^{\top} & \boldsymbol{\Sigma}_{y}
\end{array}
\right).
$$ 
%If $(\boldsymbol{\Theta},\mathbf{Y})^{\top}$ has an  augmented density representation given by \eqref{eq:ssq}, 
We  then say that   $(\boldsymbol{\Theta},\mathbf{Y})^{\top}$  is  {marginally} distributed as a $(p,q)-$variate \emph{joint projected and skew normal} with  parameters  $\boldsymbol{\mu}$,  $\boldsymbol{\Sigma}$ and $\boldsymbol{\lambda}$, i.e., $(\boldsymbol{\Theta},\mathbf{Y})^{\top}\sim \mathcal{JPSN}_{p,q}(\boldsymbol{\mu},\boldsymbol{\Sigma},\boldsymbol{\lambda})$.
%A closed from density is not available. 
Since $(\mathbf{W},\mathbf{Y})^{\top}|\mathbf{d} \sim N_{2p+q}(\boldsymbol{\mu}+( \mathbf{0}_{2p},\mbox{diag}({\boldsymbol{\lambda}}) \mathbf{d})^{\top}, \boldsymbol{\Sigma})$,   transformation of  $\mathbf{W}$ into $\boldsymbol{\Theta}$ using \eqref{eq:tranpn} implies that $(\boldsymbol{\Theta},\mathbf{Y})^{\top}$  is $\mathcal{JPSN}$ distributed; this last remark can be used to easily simulate random samples from the $\mathcal{JPSN}$.

Closure under marginalization  of the  $\mathcal{PN}$ and the $\mathcal{SSN}$ shows that any  subset of $(\boldsymbol{\Theta},\mathbf{Y})^{\top}$ is still $\mathcal{JPSN}$ distributed (see equation  \eqref{eq:ssq}) and, as limit cases,  $\boldsymbol{\Theta} \sim \mathcal{PN}_{p}(\boldsymbol{\mu}_w,\boldsymbol{\Sigma}_w)$ and $\mathbf{Y} \sim \mathcal{SSN}_{q}(\boldsymbol{\mu}_y,\boldsymbol{\Sigma}_y,\mbox{diag}(\boldsymbol{\lambda}))$. 
 %since the $\mathcal{PN}$ is closed under conditioning. 
 The flexibility of the $\mathcal{PN}$ and $\mathcal{SSN}$  are then inherited by the marginal distributions of our proposal that allows also multivariate dependence between its components.  
 {Conditional densities are not standard, but from  $(\mathbf{W},\mathbf{Y})^{\top}|\mathbf{d} \sim N_{2p+q}(\boldsymbol{\mu}+( \mathbf{0}_{2p},\mbox{diag}({\boldsymbol{\lambda}}) \mathbf{d})^{\top}, \boldsymbol{\Sigma})$ we can easily see that 
\begin{align}
\boldsymbol{\Theta}|\mathbf{y},\mathbf{d} \sim \mathcal{PN}_{p}\left(    \boldsymbol{\mu}_w+ \boldsymbol{\Sigma}_{wy}  \boldsymbol{\Sigma}_{y}^{-1}\left( \mathbf{y}-\boldsymbol{\mu}_y   -\mbox{diag}({\boldsymbol{\lambda}}) \mathbf{d}   \right) ,          \boldsymbol{\Sigma}_{w}+ \boldsymbol{\Sigma}_{wy}  \boldsymbol{\Sigma}_{y}^{-1} \boldsymbol{\Sigma}_{wy} ^{\top}  \right)
\end{align}	
and 
\begin{align}
\mathbf{Y}|\boldsymbol{\theta},\mathbf{r} \sim \mathcal{SSN}_{q}\left(    \boldsymbol{\mu}_y+\mbox{diag}({\boldsymbol{\lambda}}) \mathbf{d} + \boldsymbol{\Sigma}_{wy}^{\top}  \boldsymbol{\Sigma}_{w}^{-1}\left( \mathbf{w}-\boldsymbol{\mu}_w  \right) ,          \boldsymbol{\Sigma}_{y}+ \boldsymbol{\Sigma}_{wy}^{\top}  \boldsymbol{\Sigma}_{w}^{-1} \boldsymbol{\Sigma}_{wy}  \right).
\end{align}	
$\mathcal{JPSN}$ shapes are depicted in Figure  \ref{fig:dens}.
}

  {Notice that  $\Theta_i \perp \Theta_j$,  where $\perp$ indicates independence, iff $\mathbf{W}_i \perp \mathbf{W}_j$ and, by construction, matrix $\boldsymbol{\Sigma}_{wy}$ rules the circular-linear dependence since   iff $\mathbf{W}_i \perp Y_j$ then $ \Theta_i \perp Y_j$. Parameters $\boldsymbol{\mu}_{y}$ and  $\boldsymbol{\Sigma}_y$ are easily interpretable since 
 	%\begin{align} 
 	${\rm E}(\mathbf{Y})  =\boldsymbol{\mu}_{y}+ \boldsymbol{\lambda}  \sqrt{{2}/{\pi}}$,
 	${\rm Var}(\mathbf{Y})  =\boldsymbol{\Sigma}_{y}+ \left(1- {2}/{\pi}\right)  \mbox{diag}(\boldsymbol{\lambda} ) \mbox{diag}(\boldsymbol{\lambda} )$ and if  $[\boldsymbol{\Sigma}_y]_{j,k}=0$, where $[\boldsymbol{\Sigma}_y]_{j,k}$ indicates  the element   positioned in the $j^{th}$ row and $k^{th}$ column, then $Y_j$ and $Y_k$ are independent. $\boldsymbol{\lambda}$ controls the skewness of the linear component and $Y_i$ is normally distributed if  $\lambda_i=0$.  Parameters $\boldsymbol{\mu}_{w_i}$ and $\boldsymbol{\Sigma}_{w_i}$ determine the shape of the density of  $\Theta_i$, that is always  $\mathcal{PN}$. It is not clear how changing one of the element of  $\boldsymbol{\mu}_{w_i}$ or $\boldsymbol{\Sigma}_{w_i}$ affects the  density, but special cases exist: a circular uniform distribution is obtained with $\boldsymbol{\mu}_{w_i} = \mathbf{0}_2$ and $\boldsymbol{\Sigma}_{w_i} = d\mathbf{I}_2$,  $\boldsymbol{\mu}_{w_i} = \mathbf{0}_2$ produces an antipodal density and  $\mathcal{PN}(\boldsymbol{\mu}_{w_i} ,d\mathbf{I}_2 )$ is unimodal and symmetric. All the other statistics of the distribution that cannot be computed directly from the parameters, e.g., the circular mean, can be approximated with MC procedures.}

\section{Identifiability and Bayesian inference} \label{sec:idsolve}

Let $\mathbf{C}_w$   be a $2p\times 2p$ diagonal  matrix with $(2(i-1)+j)-th$ entry equal to $c_i>0$, where $i=1,\ldots,p $ and $j=1,2$. Then, since 
\begin{equation} \label{eq:dd}
\Theta_i = \text{ atan}^* \frac{W_{i2}}{W_{i1}}= \text{ atan}^* \frac{c_iW_{i2}}{c_iW_{i1}},
\end{equation} 
 the two random vectors $\mathbf{W}_i \sim \mathcal{N}_2 (\boldsymbol{\mu}_w,\boldsymbol{\Sigma}_w)$ and $\mathbf{C}_w\mathbf{W}_i \sim \mathcal{N}_{2p} (\mathbf{C}_w\boldsymbol{\mu}_w,\mathbf{C}_w\boldsymbol{\Sigma}_w\mathbf{C}_w)$ produce the same $\boldsymbol{\Theta}$, i.e., the $c_i$s  cancel out in equation \eqref{eq:dd}. It follows that   $\{\boldsymbol{\mu}_w,\boldsymbol{\Sigma}_w\}$ and  $\{\mathbf{C}_w\boldsymbol{\mu}_w,\mathbf{C}_w\boldsymbol{\Sigma}_w\mathbf{C}_w\}$ represent the same $\mathcal{PN}$ density  which is then  not identifiable.

%A solution is to fix the variance of each $W_{i2}$  to  1 \citep{Wang2013} but, due to the unavailability of algorithms for constrained covariance matrix estimation, this  induces a computational problem that together with the solution we are going to propose, will be discussed in Section   \ref{sec:idsolve}.

 {The $\mathcal{JPSN}$ is based on the $\mathcal{PN}$, that is also  its circular marginal distribution,  and it has  the same identification issue; for identifiability constraints on the parameters space are needed.}
%
%
%we  mentioned that the identification problem of the PN  arises by the use of the $\mbox{atan}^*$ transformation (equation \eqref{eq:tranpn}), that  produces the same circular random variable
%if $\mathbf{W}_i$ is multiplied by  a positive constant. 
%
%Since the $\mathcal{JPSN}$ is  built using the  $\mbox{atan}^*$ transformation, it suffers from the same  identification problem 
 Following and extending   \cite{Wang2013}, we  set to one the variance of each  $W_{i2}$ and from now on,  to avoid confusion, we indicate 
 $\{{\boldsymbol{\mu}},{\boldsymbol{\Sigma}},{\mathbf{W}},{\mathbf{R}} \}$
as $\{\tilde{\boldsymbol{\mu}},\tilde{\boldsymbol{\Sigma}},\tilde{\mathbf{W}},\tilde{\mathbf{R}}\}$  when such constraints are imposed;  $\boldsymbol{\lambda}$,  $\mathbf{D}$, $\boldsymbol{\mu}_y$  and ${\boldsymbol{\Sigma}}_y$ are always identified since they are related only to the linear component.
Let
\begin{equation}
\mathbf{C} = 
\left(
\begin{array}{cc}
\mathbf{C_w} & \mathbf{0}_{2p,q}  \\
 \mathbf{0}_{2p,q}^{\top} & \mathbf{I}_q
\end{array}
\right),
\end{equation}
where  $\mathbf{0}_{2p,q}^{\top} $ is a $2p\times q $ zero matrix, 
% be a $(2p+q)\times (2p+q)$ block diagonal matrix composed by the two  matrices $\mathbf{C_w}$ and $\mathbf{I}_q$,
 then  the sets $\{\tilde{\boldsymbol{\mu}}, \tilde{\boldsymbol{\Sigma}},\boldsymbol{\lambda}\}$  and  $\{{\boldsymbol{\mu}},  {\boldsymbol{\Sigma}},\boldsymbol{\lambda}\}$  
with 
\begin{align}
\boldsymbol{\mu} &=\mathbf{C}\tilde{\boldsymbol{\mu}}, \label{eq:s1}\\
\boldsymbol{\Sigma} &=\mathbf{C} \tilde{\boldsymbol{\Sigma}}\mathbf{C},\label{eq:s2}
\end{align}
produce the same $\mathcal{JPSN}$ density and   the following  relation holds:
 \begin{align} 
 f(\boldsymbol{\theta},\tilde{\mathbf{r}},\mathbf{y},\mathbf{d})  =  2^q\phi_{2p+q}((\tilde{\mathbf{w}}, \mathbf{y} )^{\top} |\tilde{\boldsymbol{\mu}}+(\mathbf{0}_{2p},\text{diag}(\boldsymbol{\lambda})\mathbf{d} )^{\top} , \tilde{\boldsymbol{\Sigma}}) \phi_q(\mathbf{d}| \mathbf{0}_q,\mathbf{I}_q ) \prod_{i=1}^p \tilde{r}_{i}= &\\
 2^q\phi_{2p+q}(\mathbf{C}(\tilde{\mathbf{w}}, \mathbf{y} )^{\top} |\mathbf{C}\tilde{\boldsymbol{\mu}}+(\mathbf{0}_{2p},\text{diag}(\boldsymbol{\lambda})\mathbf{d} )^{\top} , \mathbf{C}\tilde{\boldsymbol{\Sigma}}\mathbf{C}) \phi_q(\mathbf{d}| \mathbf{0}_q,\mathbf{I}_q ) \prod_{i=1}^p c_{i}\tilde{r}_{i}.& \label{eq:id2}
 \end{align}
 Notice that there is a one-to-one relation between sets $\{\boldsymbol{\mu},\boldsymbol{\Sigma}\}$ and   $\{\tilde{\boldsymbol{\mu}},\tilde{\boldsymbol{\Sigma}},\mathbf{C}\}$ since     $c_i= \sqrt{[\boldsymbol{\Sigma}]_{2i,2i}}$.

Due to the unavailability of MCMC algorithms for a constrained covariance matrix estimate, a computational problem arises and we show how to overcome it in the next section.

\subsection{The MCMC algorithm} \label{sec:MCMC}
Suppose to have $T$ observations drawn from a $(p,q)$-variate $\mathcal{JPSN}$, i.e., $(\boldsymbol{\Theta}_t,\mathbf{Y}_t)^{\top} \sim \mathcal{JPSN}_{p,q}(\tilde{\boldsymbol{\mu}},\tilde{\boldsymbol{\Sigma}},\boldsymbol{\lambda} )$ with $t=1,\ldots , T$. As the $\mathcal{JPSN}$ does not have a closed form density,  we introduce $\tilde{\mathbf{R}}_t=\{\tilde{R}_{ti}\}_{i=1}^p$ and  $\mathbf{D}_t$    as latent variables  and letting
$g_1(\tilde{\boldsymbol{\mu}},\tilde{\boldsymbol{\Sigma}}|\boldsymbol{\lambda})g_2(\boldsymbol{\lambda})$  be the prior distribution,  we want to evaluate the posterior of $\{ \{  \tilde{\mathbf{R}}_t\}_{t=1}^T,\{\mathbf{D}_t\}_{t=1}^T,\tilde{\boldsymbol{\mu}},\tilde{\boldsymbol{\Sigma}},\boldsymbol{\lambda}\}$ given by
\begin{equation} \label{eq:post}
\frac{\prod_{t=1}^T2^q \phi_{2p+q}((\tilde{\mathbf{w}}_t, \mathbf{y}_t )^{\top} |\tilde{\boldsymbol{\mu}} +(\mathbf{0}_{2p},\text{diag}(\boldsymbol{\lambda})\mathbf{d}_t )^{\top} , \tilde{\boldsymbol{\Sigma}}) \phi_q(\mathbf{d}_t| \mathbf{0}_q,\mathbf{I}_q ) \prod_{i=1}^p \tilde{r}_{ti} g_1(\tilde{\boldsymbol{\mu}},\tilde{\boldsymbol{\Sigma}}|\boldsymbol{\lambda})g_2(\boldsymbol{\lambda})}{Z( \{ \boldsymbol{\theta}_t,\mathbf{y}_t \}_{t=1}^T  )},
\end{equation}
where $Z( \{ \boldsymbol{\theta}_t,\mathbf{y}_t \}_{t=1}^T  )$ is the normalization constant.
Some difficulties arise in the definition of $g_1(\cdot)$  since its domain must contain  the space of constrained  nnd matrices and, to the best of our knowledge,  no priors with such domain are available.

Our proposed MCMC  algorithm starts defining  a prior $f(\boldsymbol{\mu},\boldsymbol{\Sigma}|\boldsymbol{\lambda})$ over  $\{\boldsymbol{\mu},\boldsymbol{\Sigma}\}$.
 We indicate with ${f}^*(\mathbf{C},\tilde{\boldsymbol{\mu}},\tilde{\boldsymbol{\Sigma}}|\boldsymbol{\lambda})$
the distribution over   $\{\mathbf{C},\tilde{\boldsymbol{\mu}},\tilde{\boldsymbol{\Sigma}}\}$ induced by $f(\boldsymbol{\mu},\boldsymbol{\Sigma}|\boldsymbol{\lambda})$    and  we  define  $g_1(\tilde{\boldsymbol{\mu}},\tilde{\boldsymbol{\Sigma}}|\boldsymbol{\lambda})$  as 
\begin{equation} \label{eq:priorg2}
g_1(\tilde{\boldsymbol{\mu}},\tilde{\boldsymbol{\Sigma}}|\boldsymbol{\lambda}) = \int_{\mathbb{R}^+}\ldots \int_{\mathbb{R}^+}   {f}^*(\mathbf{C},\tilde{\boldsymbol{\mu}},\tilde{\boldsymbol{\Sigma}}|\boldsymbol{\lambda})   d c_1  \ldots d c_p.
\end{equation}
%This way to define $g_1()$, and with  appropriate choices of $f()$, let us able to define an MCMC algorithm really easy to implement. 		\\

Then, using   \eqref{eq:id2} and \eqref{eq:priorg2} we can  write  \eqref{eq:post}  as 
\begin{align} 
\int_{\mathbb{R}^+}\ldots \int_{\mathbb{R}^+}  & \prod_{t=1}^T 2^q  \phi_{2p+q}(\mathbf{C}(\tilde{\mathbf{w}}_t, \mathbf{y}_t )^{\top} |\mathbf{C}\tilde{\boldsymbol{\mu}} +(\mathbf{0}_{2p},\text{diag}(\boldsymbol{\lambda})\mathbf{d} )^{\top} , \mathbf{C}\tilde{\boldsymbol{\Sigma}}\mathbf{C}) \times \\ 
&   \frac{\phi_q(\mathbf{d}_t| \mathbf{0}_q,\mathbf{I}_q )  
	\prod_{i=1}^p  c_i\tilde{r}_{ti}  {f}^*(\mathbf{C},\tilde{\boldsymbol{\mu}},\tilde{\boldsymbol{\Sigma}}|\boldsymbol{\lambda}) g_2(\boldsymbol{\lambda}) }{Z( \{ \boldsymbol{\theta}_t,\mathbf{y}_t \}_{t=1}^T  )} d c_1  \ldots d c_p,\label{eq:asd}
\end{align}
and if we transform  $\{\mathbf{C},\{\tilde{\mathbf{R}}_t  \}_{t=1}^T,\tilde{\boldsymbol{\mu}},\tilde{\boldsymbol{\Sigma}} \}$ into  $\{\{\mathbf{R}_t  \}_{t=1}^T,\boldsymbol{\mu}, \boldsymbol{\Sigma}  \}$,  the integrand of equation  \eqref{eq:asd} becomes 
\begin{equation} \label{eq:ww2}
\frac{\prod_{t=1}^T\phi_{2p+q}((\mathbf{w}_t, \mathbf{y}_t)^{\top}   |{\boldsymbol{\mu}} +(\mathbf{0}_{2p},\text{diag}(\boldsymbol{\lambda})\mathbf{d}_t )^{\top}, {\boldsymbol{\Sigma}}) \phi_q(\mathbf{d}_t| \mathbf{0}_q,\mathbf{I}_q )   \prod_{i=1}^p r_{ti} f(\boldsymbol{\mu},\boldsymbol{\Sigma}|\boldsymbol{\lambda}) g_2(\boldsymbol{\lambda})}{Z\left(\{  \boldsymbol{\theta}_t,\mathbf{y}_t\}_{t=1}^T\right)}.
\end{equation}

Then,  relying on standard  MC integration rules \cite[see for example][]{brooks2011,Robert2005},  
a set of $B$ draws from  \eqref{eq:post} is  obtained by taking $B$ samples of $\{\{\mathbf{R}_t  \}_{t=1}^T,\{{\mathbf{D}}_t  \}_{t=1}^T,\boldsymbol{\mu}, \boldsymbol{\Sigma},\boldsymbol{\lambda} \}$ from \eqref{eq:ww2} and  transforming them to $\{\{\tilde{\mathbf{R}}_t  \}_{t=1}^T,\{{\mathbf{D}}_t  \}_{t=1}^T,\tilde{\boldsymbol{\mu}},\tilde{\boldsymbol{\Sigma}} ,\boldsymbol{\lambda}\}$.

In a schematic way our proposal is 
\begin{itemize}
	\item to define a prior over  $\{\boldsymbol{\mu},\boldsymbol{\Sigma},\boldsymbol{\lambda}\}$ that induces a prior   $g_1(\cdot)$  (see equation \eqref{eq:priorg2});
%	\item to  base the steps of the  MCMC algorithm (showed below) on density  \eqref{eq:ww2}, that is defined using the non constrained nnd matrix;
	\item to obtain a set of samples of  $	\{\{\mathbf{R}_t  \}_{t=1}^T,\{{\mathbf{D}}_t  \}_{t=1}^T,\boldsymbol{\mu}, \boldsymbol{\Sigma},\boldsymbol{\lambda}\}$ from distribution \eqref{eq:ww2};
	\item to transform the posterior samples of  $\{\{\mathbf{R}_t  \}_{t=1}^T,\boldsymbol{\mu}, \boldsymbol{\Sigma}\}$ into  $\{\{\tilde{\mathbf{R}}_t  \}_{t=1}^T,\tilde{\boldsymbol{\mu}},\tilde{\boldsymbol{\Sigma}} \}$ after the model fitting.
\end{itemize}
The resulting posterior samples  are from  the distribution of interest (equation \eqref{eq:post}).  {The proposed MCMC algorithm  can be used with  the $\mathcal{JPSN}$, the univariate projected normal ($q=0$ and $p=1$), the multivariate projected normal ($q=0$) and also  with the proposal of \citep{Stumpfhause2016}, i.e., a distribution defined over the $K$-dimensional sphere,  since all of them share the same identification problem.}

 {There are no restrictions on the choice of $g_1(\cdot)$ and $g_2(\cdot)$ but,   as shown in Appendix \ref{sec:app2}, if  ease of implementation and conjugate priors are required,    a normal inverse-Wishart ($\mathcal{NIW}$)  can be used  for $\{\boldsymbol{\mu},\boldsymbol{\Sigma}\}$  and a normal  for $\lambda$; these are the ones we use in the examples of Section 	\ref{sec:ex}.}  {Regardless of the priors chosen, the updates}  of $\mathbf{D}_t$ and $R_{ti}$ can be done using Gibbs steps.

\section{Examples} \label{sec:ex}

\subsection{Synthetic data} \label{sec:sim}

\newsavebox{\AAA}
\savebox{\AAA}{$[\tilde{\boldsymbol{\mu}}_k]_{1}$}
\newsavebox{\BBB}
\savebox{\BBB}{$[\tilde{\boldsymbol{\Sigma}}_k]_{1,1}$}
\newsavebox{\CCC}
\savebox{\CCC}{$[\tilde{\boldsymbol{\Sigma}}_k]_{1,2}$}

\begin{figure}[t!]
	\centering 
	\captionsetup[subfigure]{labelformat=empty}
	\subfloat{\raisebox{+0.68in}{\rotatebox[origin=t]{90}{Example 1}}}
	{\subfloat{\includegraphics[trim= {0.7cm 1.cm 0.7cm 1.cm},clip,scale=0.22]{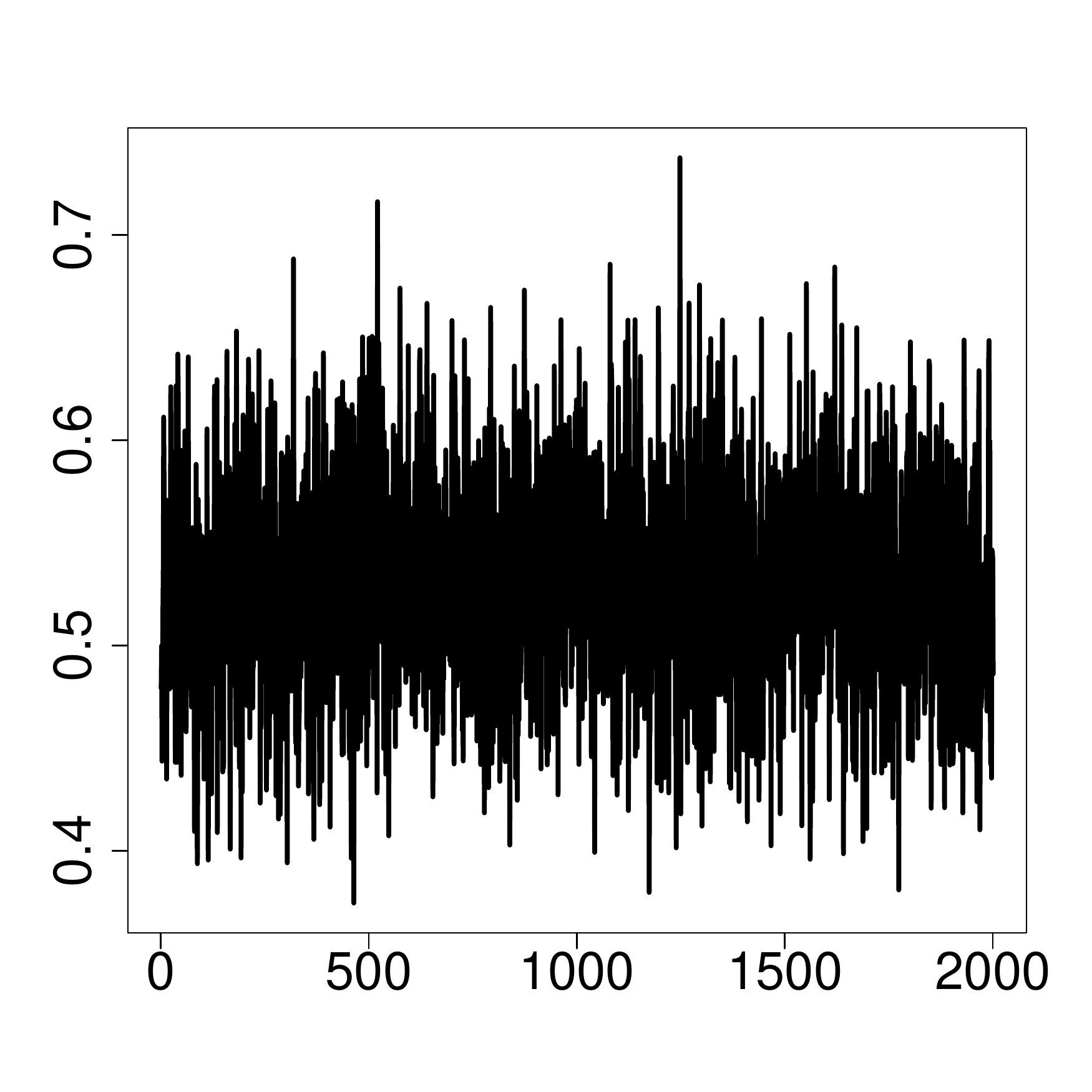}}}
	{\subfloat{\includegraphics[trim= {0.7cm 1.cm 0.7cm 1.cm},clip,scale=0.22]{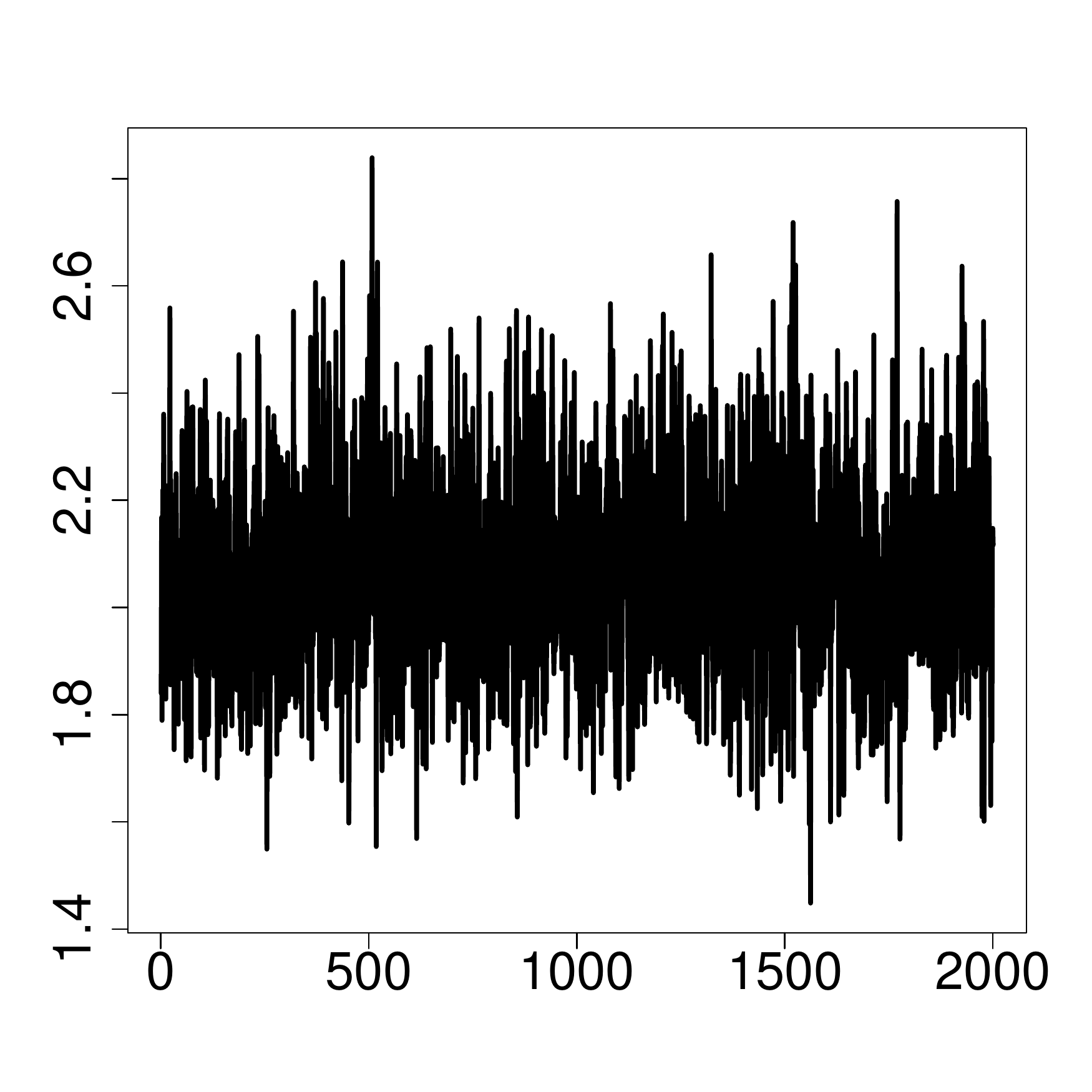}}}
	{\subfloat{\includegraphics[trim= {0.7cm 1.cm 0.7cm 1.cm},clip,scale=0.22]{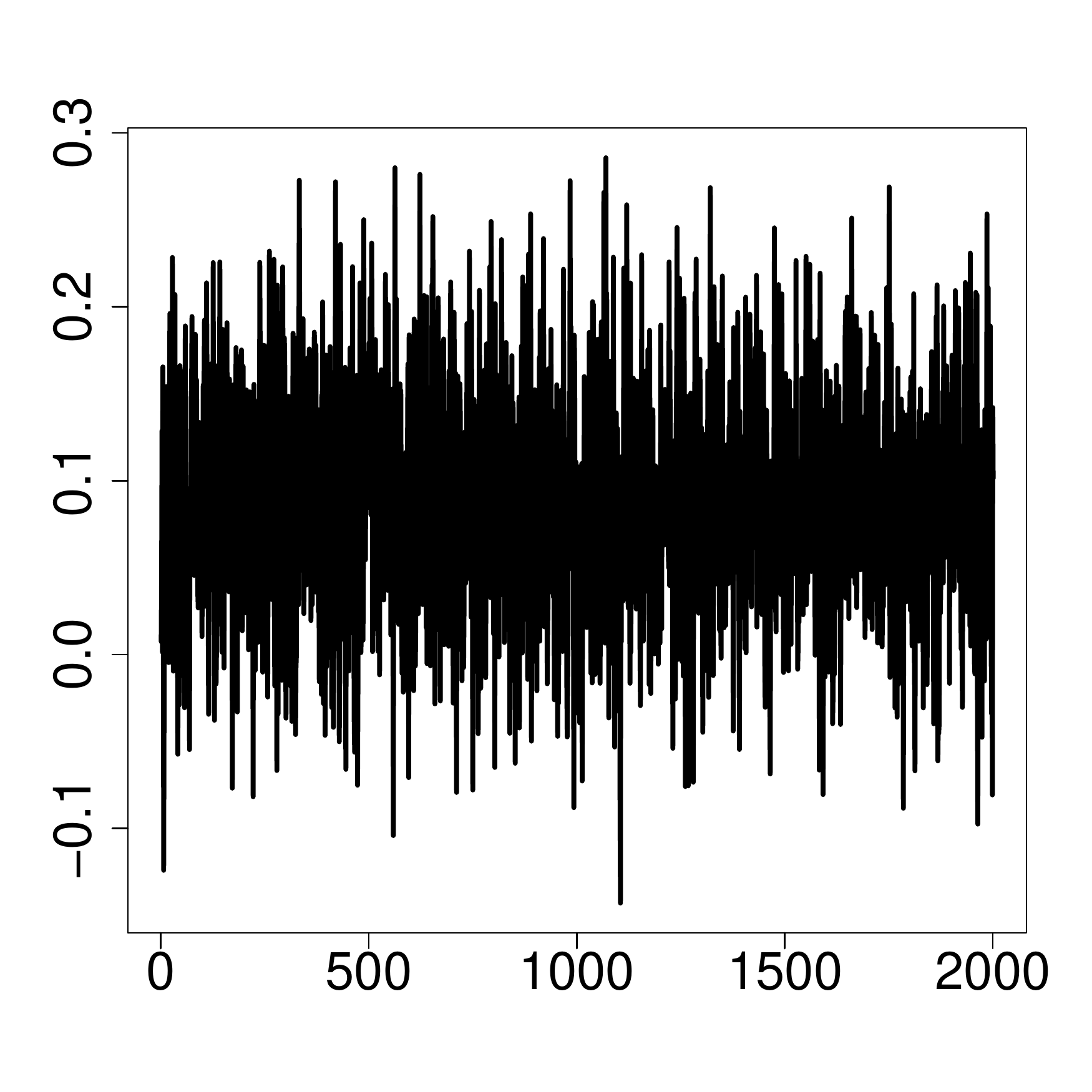}}}\\
	\subfloat{\raisebox{+0.68in}{\rotatebox[origin=t]{90}{Example 2}}}
	{\subfloat{\includegraphics[trim= {0.7cm 1.cm 0.7cm 1.cm},clip,scale=0.22]{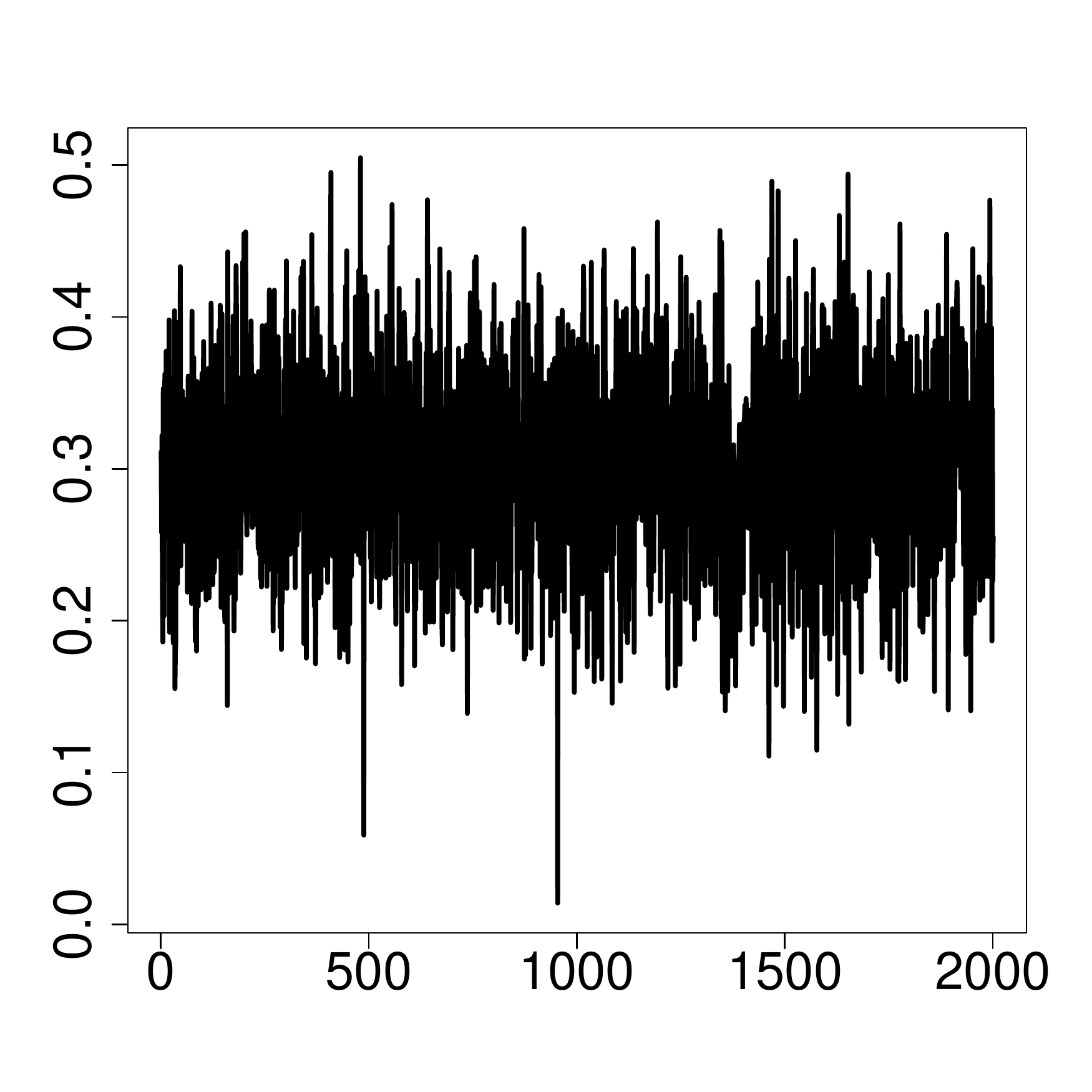}}}
	{\subfloat{\includegraphics[trim= {0.7cm 1.cm 0.7cm 1.cm},clip,scale=0.22]{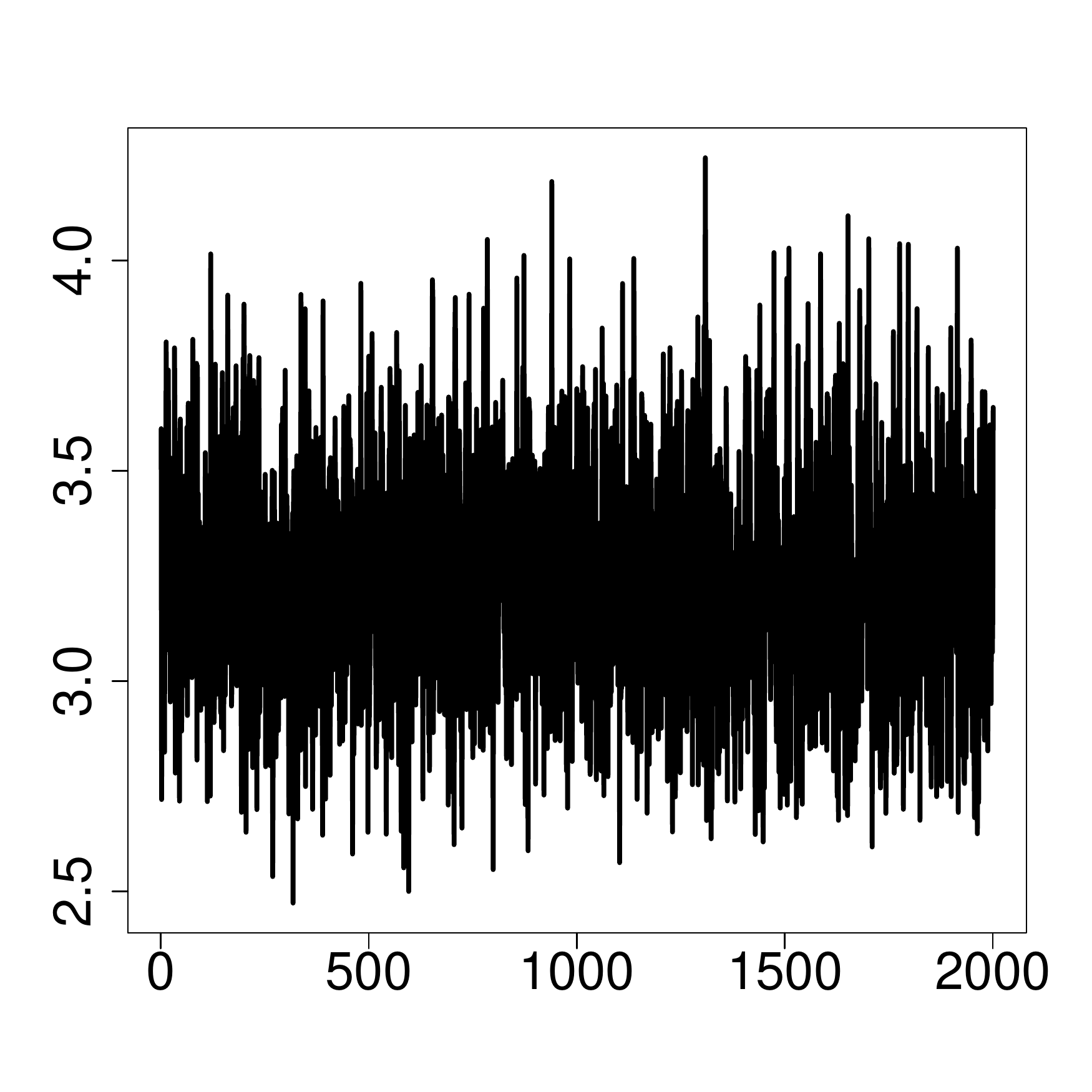}}}
	{\subfloat{\includegraphics[trim= {0.7cm 1.cm 0.7cm 1.cm},clip,scale=0.22]{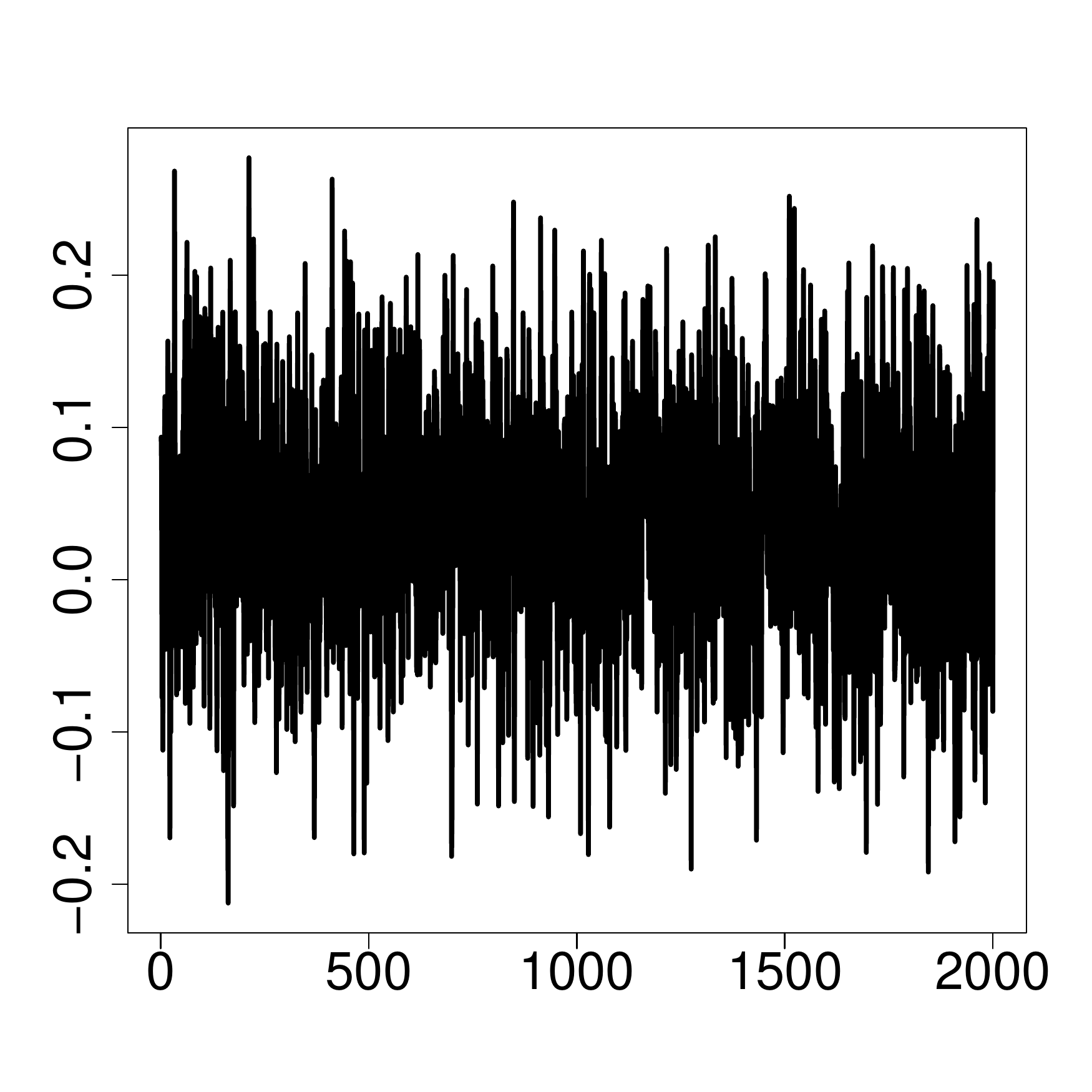}}}\\
	\subfloat{\raisebox{+0.68in}{\rotatebox[origin=t]{90}{Example 3}}}
	{\subfloat[ \usebox{\AAA} ]{\includegraphics[trim= {0.7cm 1.cm 0.7cm 1.cm},clip,scale=0.22]{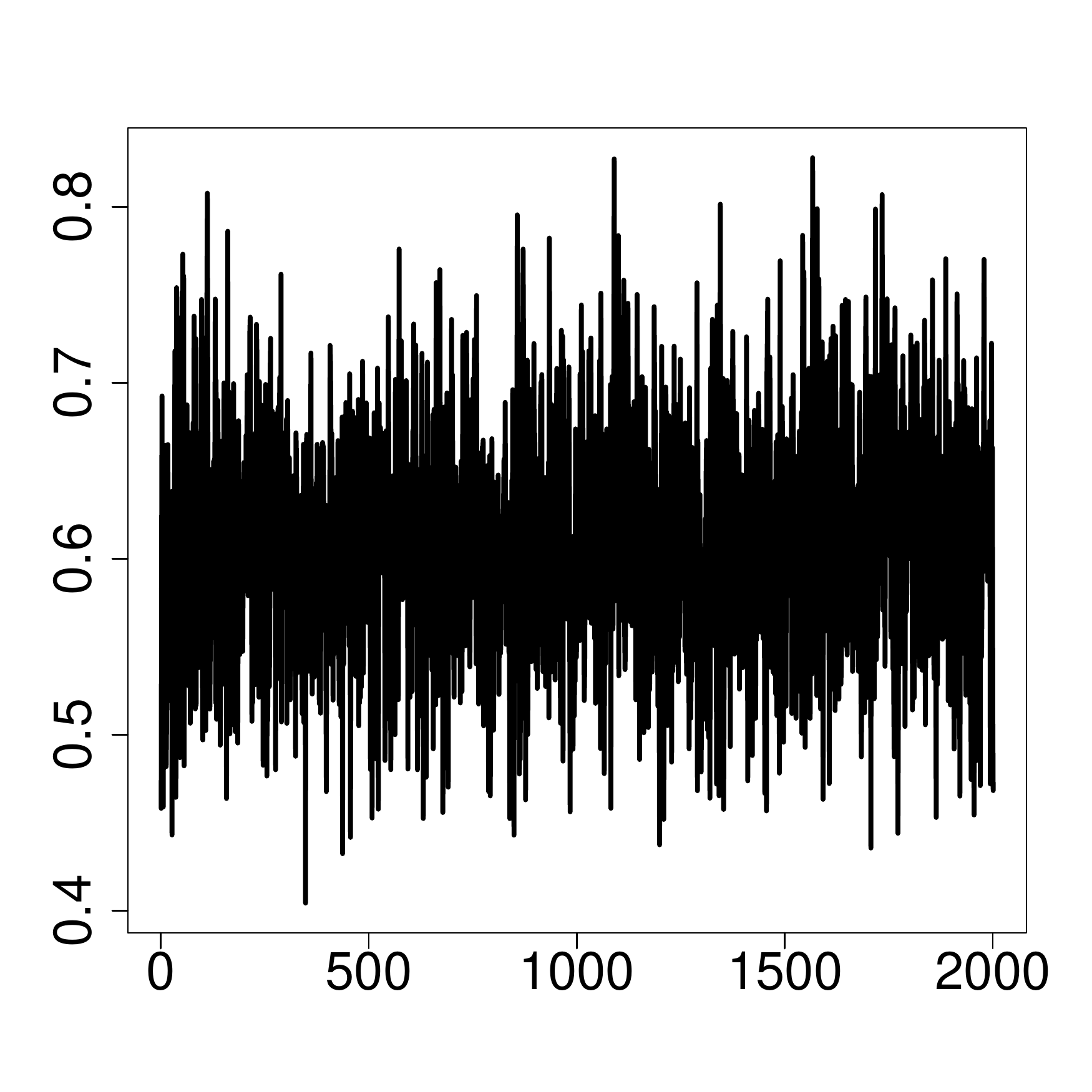}}}
	{\subfloat[\usebox{\BBB}]{\includegraphics[trim= {0.7cm 1.cm 0.7cm 1.cm},clip,scale=0.22]{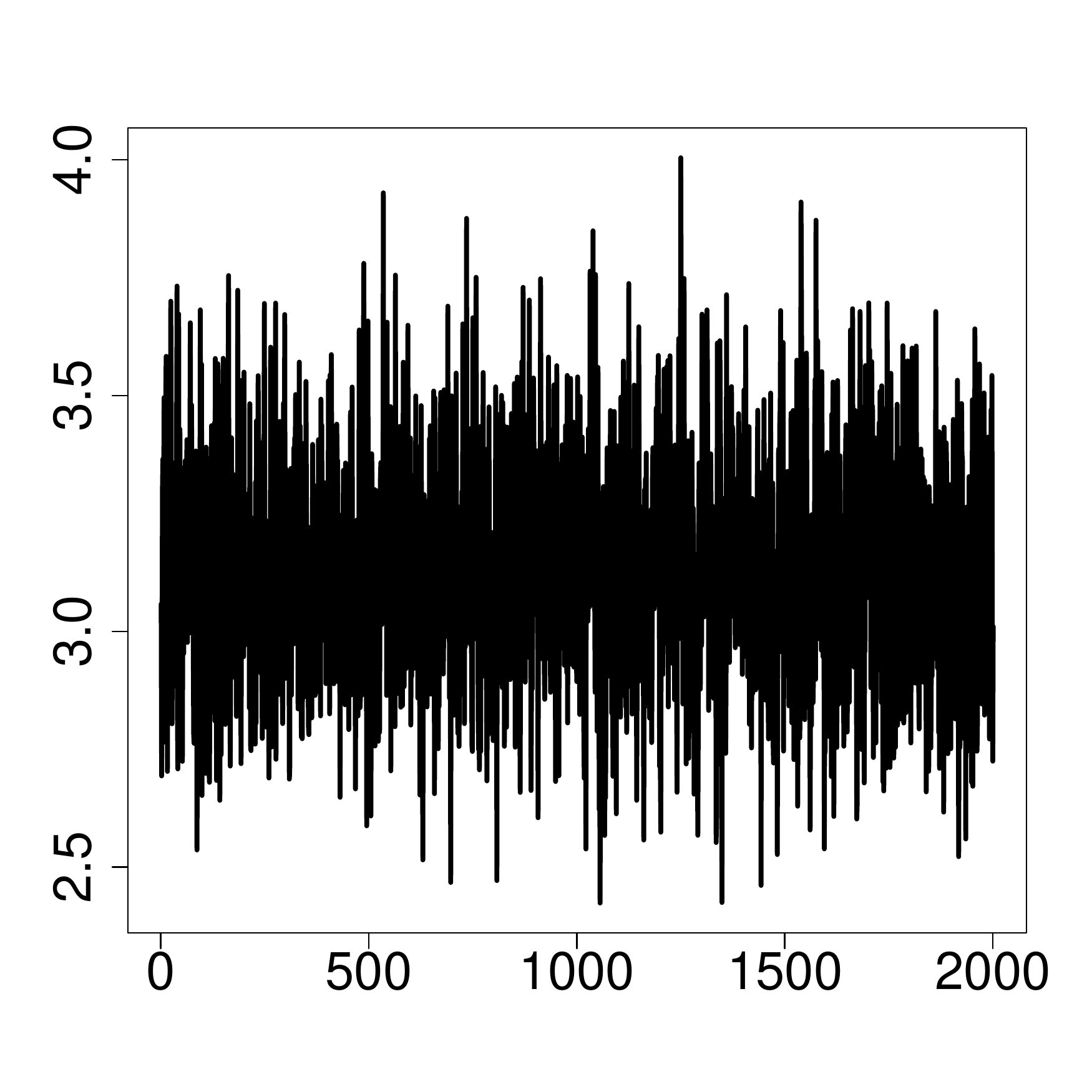}}}
	{\subfloat[\usebox{\CCC}]{\includegraphics[trim= {0.7cm 1.cm 0.7cm 1.cm},clip,scale=0.22]{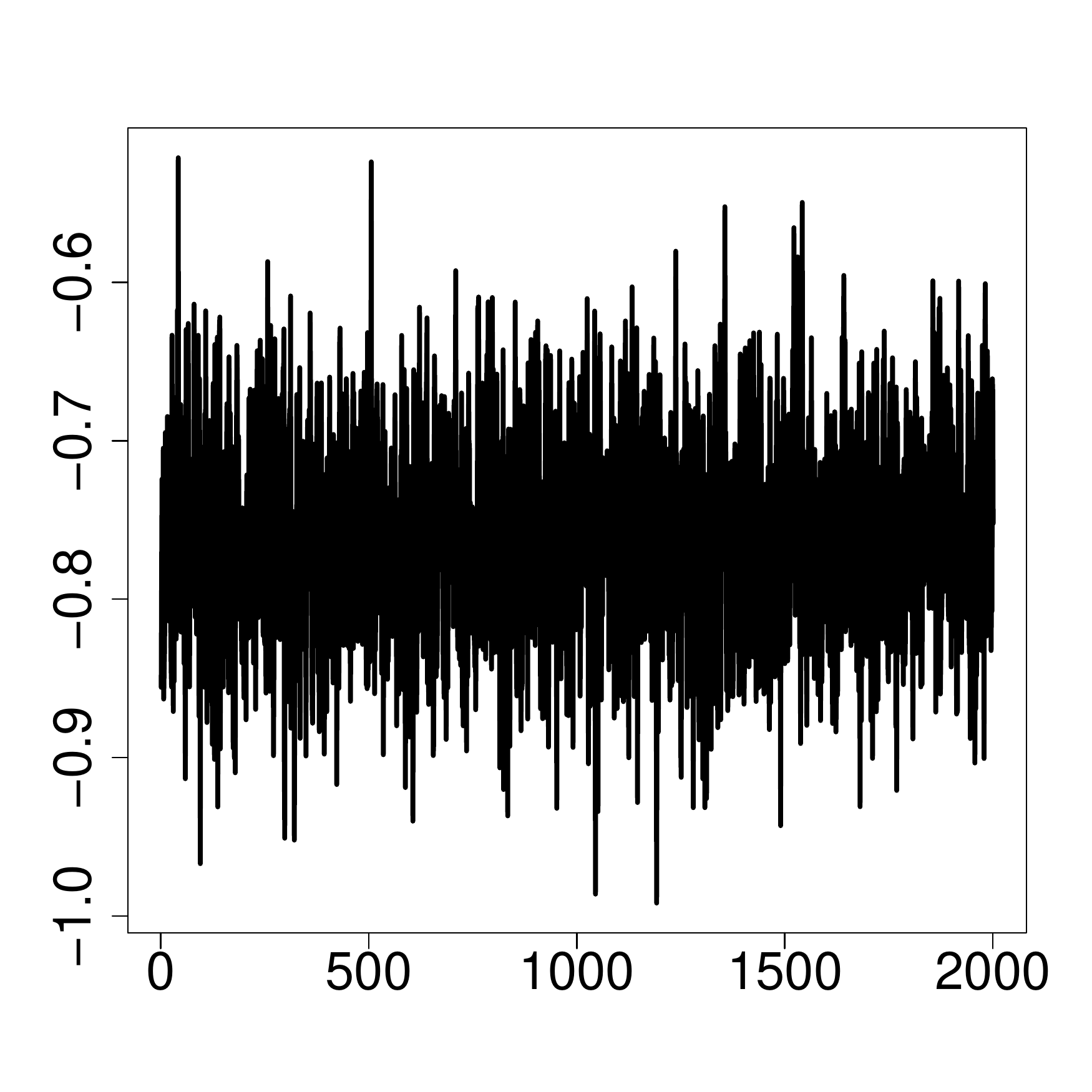}}}
	\caption{Simulated examples - trace plots of   parameters   $[\tilde{\boldsymbol{\mu}}_k]_{1}$, $[\tilde{\boldsymbol{\Sigma}}_k]_{1,1}$ and $[\tilde{\boldsymbol{\Sigma}}_k]_{1,2}$ (columns)  in the three examples (rows)  }  \label{fig:vvv}
\end{figure}

\begin{table}[t!]
	\centering
	\begin{tabular}{c|ccccc}
		\hline \hline
		&& Example\\
		& k=1 & k=2 & k=3\\\hline 
		$ [\hat{\tilde{\boldsymbol{\mu}}}_k]_{1} $ & 0.529 & 0.301 & 0.608\\
		CI&	(0.428 0.638)& (0.179 0.427) & (0.480 0.738)\\
		$[\hat{\tilde{\boldsymbol{\mu}}}_k]_{2}$  & -0.943 &0.207  & 0.5\\
		CI& (-1.029 -0.858)& (0.137 0.280) & (0.428 0.576)\\
		$[\hat{\tilde{\boldsymbol{\mu}}}_k]_{3}$  & -0.112  & -0.02 & 0.007\\
		CI& (-0.145 -0.080)& (-0.068  0.029) & (-0.023  0.036)\\
		$[\hat{\tilde{\boldsymbol{\mu}}}_k]_{4}$  & 0.095&  0.111 & 0.503\\
		CI& (0.020 0.163)& (0.041 0.178) & (0.429 0.576)\\
		$[\hat{\tilde{\boldsymbol{\mu}}}_k]_{5}$  &-5.095  &  -4.765& 4.705 \\
		CI& (-5.415 -4.780)& (-4.925 -4.596) & (4.489 5.105)\\
		$\hat{\lambda}_{k}$  &  -4.941  &  4.872 & 6.237 \\
		CI& (-5.320 -4.561)& (4.630 5.135) & (5.920 6.556)\\ \hline \hline
	\end{tabular}
	\caption{Simulated examples - posterior mean  estimates ( $\hat{}$ ) and 95\% credible intervals (CI)  of $\tilde{\boldsymbol{\mu}}_k$ and ${\lambda}_k$.} \label{tab:simes}
\end{table}

\begin{table}
		\centering
	\small{
		\begin{tabular}{c|ccccc}
			\hline \hline
				&j=1&j=2&j=3&j=4&j=5  \\\hline 
			$[\hat{\tilde{\boldsymbol{\Sigma}}}_1]_{1,j}$ &2.062& 0.085 & -0.031 & 0.011 & -0.001\\
			CI&(1.726 2.462)&(-0.040  0.214)&(-0.083  0.021)&(-0.098  0.121)&(-0.287  0.278)\\
			$[\hat{\tilde{\boldsymbol{\Sigma}}}_1]_{2,j}$ & $\cdot$ & 1 &  -0.008 & -0.009 &  0.149\\
			CI&($\cdot$  $\cdot$)&(1 1)&(-0.046  0.028)&(-0.094  0.075)&(-0.060  0.368)\\
			$[\hat{\tilde{\boldsymbol{\Sigma}}}_1]_{3,j}$ &  $\cdot$&   $\cdot$&  0.201& 0.002 & 0.007 \\
			CI&($\cdot$  $\cdot$)&($\cdot$  $\cdot$)&(0.168 0.237)&(-0.039  0.041)&(-0.079  0.097)\\
			$[\hat{\tilde{\boldsymbol{\Sigma}}}_1]_{4,j}$ & $\cdot$ &  $\cdot$ &  $\cdot$ & 1 & 0.047\\
			CI&($\cdot$  $\cdot$)&($\cdot$  $\cdot$)&($\cdot$  $\cdot$)&(1 1)&(-0.144  0.248)\\
			$[\hat{\tilde{\boldsymbol{\Sigma}}}_1]_{5,j}$ & $\cdot$ &  $\cdot$ &  $\cdot$ & $\cdot$  &  2.151\\
			CI&($\cdot$  $\cdot$)&($\cdot$  $\cdot$)&($\cdot$  $\cdot$)&($\cdot$  $\cdot$)&(1.532 2.908)\\ \hline \hline
					$[\hat{\tilde{\boldsymbol{\Sigma}}}_2]_{1,j}$ &3.23& 0.04 & 0.581 & 0.862 & 0.882\\
					CI&(2.719 3.793)&(-0.112  0.191)&(0.467 0.713)&(0.722 1.015)&(0.619 1.151)\\
					$[\hat{\tilde{\boldsymbol{\Sigma}}}_2]_{2,j}$ & $\cdot$ & 1 & -0.293 & 0.429 & 0.419\\
					CI&($\cdot$  $\cdot$)&(1 1)&(-0.354 -0.235)&(0.361 0.492)&(0.284 0.557)\\
					$[\hat{\tilde{\boldsymbol{\Sigma}}}_2]_{3,j}$ &  $\cdot$&   $\cdot$& 0.521 & 0.034 &-0.223 \\
					CI&($\cdot$  $\cdot$)&($\cdot$  $\cdot$)&(0.438 0.617)&(-0.024  0.095)&(-0.333 -0.120)\\
					$[\hat{\tilde{\boldsymbol{\Sigma}}}_2]_{4,j}$ & $\cdot$ &  $\cdot$ &  $\cdot$ & 1 & 0.496\\
					CI&($\cdot$  $\cdot$)&($\cdot$  $\cdot$)&($\cdot$  $\cdot$)&(1 1)&(0.354 0.645)\\
					$[\hat{\tilde{\boldsymbol{\Sigma}}}_2]_{5,j}$ & $\cdot$ &  $\cdot$ &  $\cdot$ & $\cdot$  &1.001 \\
					CI&($\cdot$  $\cdot$)&($\cdot$  $\cdot$)&($\cdot$  $\cdot$)&($\cdot$  $\cdot$)&(0.737 1.316)\\ 
						\hline \hline
						$[\hat{\tilde{\boldsymbol{\Sigma}}}_3]_{1,j}$ &3.129&  -0.762& 0.38 &  0.623 & 0.834\\
						CI&(2.684 3.620)&(-0.889 -0.633)&(0.309 0.455)&(0.469 0.775)&(0.566 1.132)\\
						$[\hat{\tilde{\boldsymbol{\Sigma}}}_3]_{2,j}$ & $\cdot$ & 1 & 0.207  & 0.389  &-0.176 \\
						CI&($\cdot$  $\cdot$)&(1 1)&(0.177 0.238)&(0.319 0.458)&(-0.325 -0.021)\\
						$[\hat{\tilde{\boldsymbol{\Sigma}}}_3]_{3,j}$ &  $\cdot$&   $\cdot$&  0.189& 0.223 & 0.17\\
						CI&($\cdot$  $\cdot$)&($\cdot$  $\cdot$)&(0.164 0.216)&(0.193 0.255)&(0.108 0.240)\\
						$[\hat{\tilde{\boldsymbol{\Sigma}}}_3]_{4,j}$ & $\cdot$ &  $\cdot$ &  $\cdot$ & 1 & -0.337\\
						CI&($\cdot$  $\cdot$)&($\cdot$  $\cdot$)&($\cdot$  $\cdot$)&(1 1)&(-0.493 -0.188)\\
						$[\hat{\tilde{\boldsymbol{\Sigma}}}_3]_{5,j}$ & $\cdot$ &  $\cdot$ &  $\cdot$ & $\cdot$  &0.875 \\
						CI&($\cdot$  $\cdot$)&($\cdot$  $\cdot$)&($\cdot$  $\cdot$)&($\cdot$  $\cdot$)&(0.620 1.165)\\ \hline \hline
				\end{tabular}}
				\caption{Simulated examples - posterior mean  estimates ( $\hat{}$ ) and 95\% credible intervals (CI) of $\tilde{\boldsymbol{\Sigma}}_{k}$.  }   \label{tab:simes4}
			\end{table}

The aim of these simulated examples is to prove that the proposed MCMC algorithm is able to retrieve the parameters used to simulate the data and to solve the identification problem. 
We simulate  3 datasets with $p=2$, $q=1$, i.e., two circular variables  and 1 linear,   $T=1000$  and parameters 
\begin{equation}
\tilde{\boldsymbol{\mu}}_1 = \left[
\begin{array}{r}
0.5\\-1.0\\-0.1\\0.1\\	-5.0
\end{array}
\right], \quad
\tilde{\boldsymbol{\mu}}_2 = \left[
\begin{array}{r}
0.2\\0.2\\0.0\\0.1\\	-5.0
\end{array}
\right], \quad
\tilde{\boldsymbol{\mu}}_3 = \left[
\begin{array}{r}
0.5\\0.5\\0.0\\0.5\\5.0
\end{array}
\right],
\end{equation}
$\lambda_{1}=-5$, $\lambda_{2}=5$, $\lambda_{3}=6$,
\begin{equation}
\tilde{\boldsymbol{\Sigma}}_1 = 
\left[
\begin{array}{rrrrr}
2 &   0 & 0.0&    0 &   0 \\
0  &  1 & 0.0  &  0  &  0\\
0  &  0 & 0.2  &  0  &  0\\
0  &  0 & 0.0  &  1  &  0\\
0  &  0&  0.0  &  0&    2
\end{array}
\right], \quad \tilde{\boldsymbol{\Sigma}}_2 = 
\left[
\begin{array}{rrrrr}
3.000 & 0.000 & 0.551& 0.779 & 0.857\\
0.000 & 1.000 &-0.318& 0.450 & 0.495\\
0.551 &-0.318 & 0.500& 0.000 &-0.318\\
0.779 & 0.450&  0.000& 1.000&  0.450\\
0.857  &0.495& -0.318 &0.450&  1.000
\end{array}
\right],
\end{equation}

\begin{equation}
\tilde{\boldsymbol{\Sigma}}_3 = 
\left[
\begin{array}{rrrrr}
3.000& -0.783 &0.377 & 0.684 & 0.781\\
-0.783 & 1.000& 0.214&  0.335 &-0.092\\
0.377 & 0.214 &0.200 & 0.231&  0.209\\
0.684 & 0.335 &0.231 & 1.000 &-0.382\\
0.781 &-0.092 &0.209 &-0.382 & 1.000
\end{array}
\right]. 
\end{equation}
The marginal bivariate densities  are plotted in Figure \ref{fig:dens}.   We chose the parameters so to have independent (first example) and dependent variables (second and third), highly skew linear densities and, at least, one bimodal circular marginal for each example.

In the three examples  inference is carried out considering  40000 iterations, burnin 30000, thin 5 and by taking  2000 posterior samples. As  prior distributions  we choose  $\boldsymbol{\mu}_k,\boldsymbol{\Sigma}_k \sim \mathcal{NIW} \left( \mathbf{0}_{5}, 0.001, 15, \mathbf{I}_{5}  \right)$ and ${\lambda}_k \sim \mathcal{N}_1\left( 0, 100  \right)$, that are standard weak informative priors. From Tables \ref{tab:simes} and \ref{tab:simes4} we  see that, with the exception of $[\hat{\tilde{\boldsymbol{\mu}}}_2]_{5}$, all true values   are inside the associated 95\% credible intervals (CIs), proving that our algorithm is able to estimate the $\mathcal{JPSN}$ parameters.   To further corroborate the validity of the proposed MCMC scheme in solving the identification problem, in  Figure \ref{fig:vvv} we show, as  examples, the trace plots of parameters $[\tilde{\boldsymbol{\mu}}_k]_{1}$, $[\tilde{\boldsymbol{\Sigma}}_k]_{1,1}$ and $[\tilde{\boldsymbol{\Sigma}}_k]_{1,2}$.  
These chains have reached their stationary distributions (we also checked it by using  the  \textsf{R} package \texttt{coda} \citep{coda}) with weak informative priors; this shows that the identification problem is no more relevant.  

\subsection{Zebras movements  example} \label{sec:real2}

\begin{figure}[t!]
	\centering
	\captionsetup[subfigure]{labelformat=empty}
	\subfloat{\raisebox{+0.48in}{\rotatebox[origin=t]{90}{Circular}}}
	{\subfloat{\includegraphics[trim= -5 20 50 20,scale=0.26]{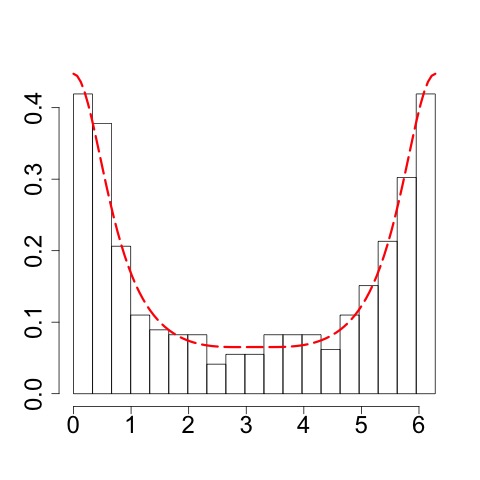}}}
	{\subfloat{\includegraphics[trim= -5 20 50 20,scale=0.26]{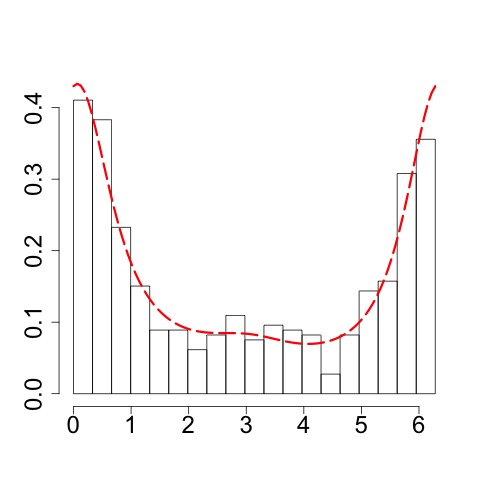}}}
	{\subfloat{\includegraphics[trim= -5 20 50 20,scale=0.26]{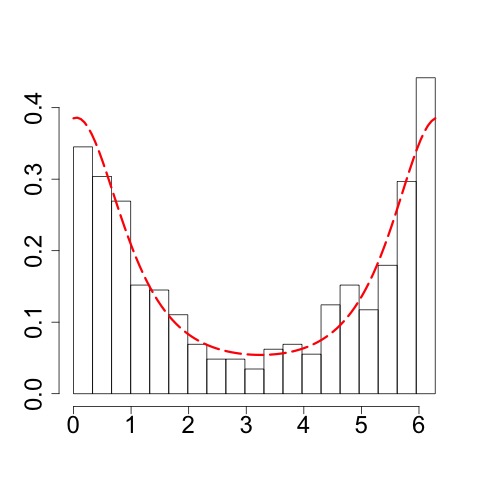}}}
	{\subfloat{\includegraphics[trim= -5 20 50 20,scale=0.26]{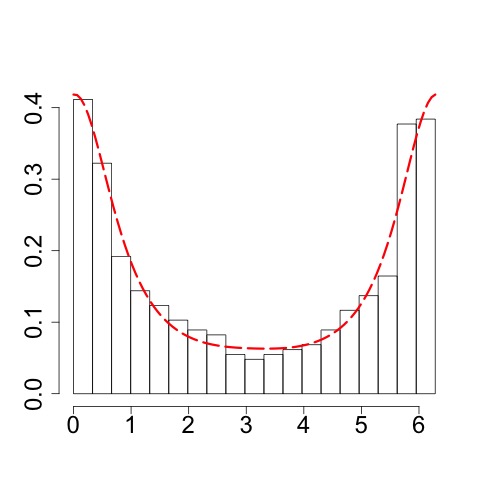}}}\\
	\subfloat{\raisebox{+0.48in}{\rotatebox[origin=t]{90}{Linear}}}
	{\subfloat[First zebra]{\includegraphics[trim= -5 20 50 20,scale=0.26]{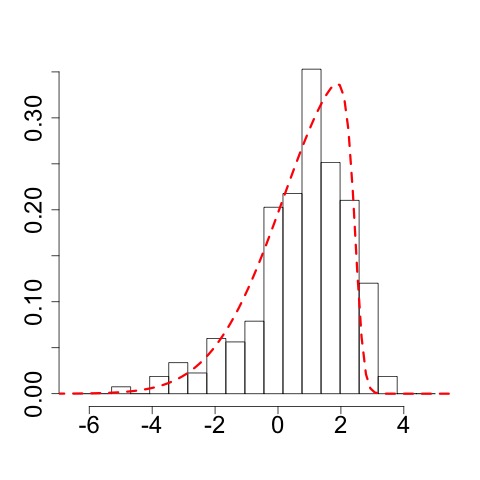}}}
	{\subfloat[Second zebra]{\includegraphics[trim= -5 20 50 20,scale=0.26]{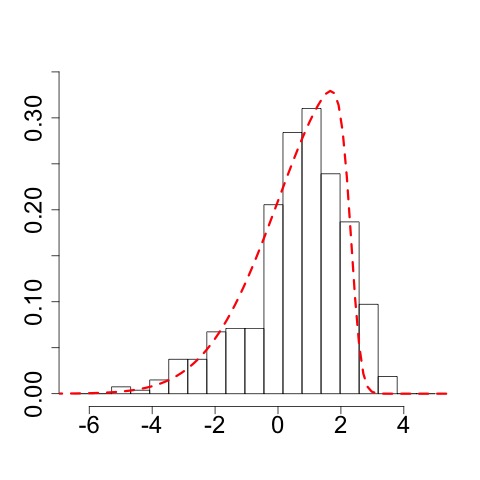}}}
	{\subfloat[Third zebra]{\includegraphics[trim= -5 20 50 20,scale=0.26]{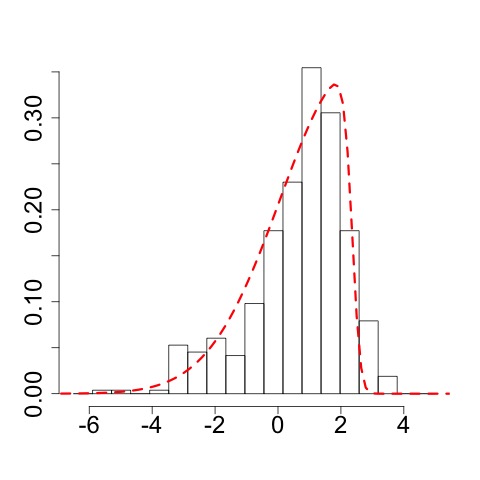}}}
	{\subfloat[Fourth zebra]{\includegraphics[trim= -5 20 50 20,scale=0.26]{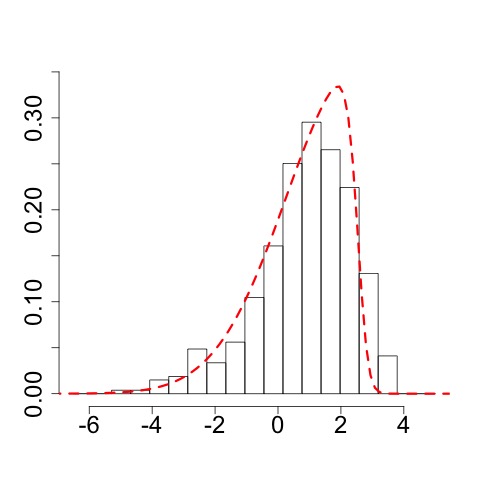}}}
	\caption{Zebras  movement example -  histograms of the observed data and posterior marginal densities of  turning-angles (first row)  and the logarithm of step-lengths (second row). } \label{fig:ESTDlin2}
\end{figure}

\begin{figure}[t]
	\centering
	{{\includegraphics[trim= 35 35 35 35,scale=0.50]{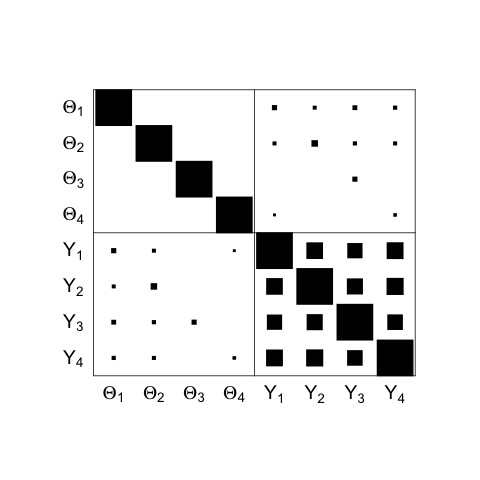}}}
	\caption{Zebras movement example - dependence matrix computed using  equations \eqref{eq:112} (circular-circular), \eqref{eq:111} (circular-linear) and the Pearson's correlation coefficient (linear-linear). The size of the square is proportional to the posterior mean value. All  values are positive.} \label{fig:Cor2}
\end{figure}

 {In this section we estimate the $\mathcal{JPSN}$ parameters on  an animal movement  dataset  taken from the movebank repository (\url{www.movebank.org}). Our aim is to show that the $\mathcal{JPSN}$  can give    information  on the  dependence of poly-cylindrical observations.  }
Seven zebras are jointly observed in Botswana (Africa)  between the Okavango Delta  and the Makgadikgadi Pans, and their hourly positions are recorded with GPS devices during the years 2007-2009 \citep{Bartlam2013}.  In the observational period the zebras migrate from  the dry season habitat, that is the  Okavango Delta, to the  rainy season habitat, that is  the Makgadikgadi Pans. We  select data from 4 zebras, observed between the 18 of November 2008 and the 18 of February 2009, when they have ended the migration.
For each animal we compute the turning-angles and  logarithm of  step-lengths, having then poly-cylindrical observations composed of
 four circular and four linear variables.  {It is out of the scope of this work to introduce complex models based on the $\mathcal{JPSN}$, that are left to future developments,  and we assume that  observations  are  independent and identical distributed. For this reason,  to mitigate temporal dependence  we use data  five hours apart, having then 442 observations for parameters  estimate. Using the Pearson's coefficient and the    circular-circular correlation of \cite{Jammalamadaka1988}, i.e., 
 	\begin{equation}
 	\rho_{(\Theta_i,\Theta_{i^{\prime}})} =    \frac{{\rm E}(\sin(\Theta_i-\Theta_i^*)\sin(\Theta_{i^{\prime}}-\Theta_{i^{\prime}}^*)    )}{\sqrt{ {\rm E}(\sin^2(\Theta_i-\Theta_i^*)) {\rm E}(\sin^2(\Theta_{i^{\prime}}-\Theta_{i^{\prime}}^*)   )     }}  \in [-1,1]
 	,\label{eq:112}
 	\end{equation}
 	where $\Theta_i^*$ and $\Theta_{i^{\prime}}^*$ are two circular variables  distributed, respectively, as $\Theta_i$ and $\Theta_{i^{\prime}}$,  for the subset of data used   all the  autocorrelations have values lower than 0.05. } The  histograms of the data  can be seen in Figure \ref{fig:ESTDlin2}.

%
%
%Indeed, important features are the circular mean and concentration  of $\Theta_{i}$,  i.e.  circular counterparts of the mean and precision of a linear variable,  respectively equal to
%$
%\alpha_{i} = \mbox{atan}^* {{\rm E}(\sin \Theta_{i} )}/{{\rm E}(\cos \Theta_{i} )}
%$
%and 
%$
%\zeta_{i} = \sqrt{{\rm E}^2(\sin \Theta_{i} )+{\rm E}^2(\cos \Theta_{i} )},
%$
%and the measures of circular-circular and circular-linear association as
%the circular-circular correlation of  \cite{Jammalamadaka1988}, i.e.
%\begin{equation}
%\rho_{(\Theta_i,\Theta_{i^{\top}})} =    \frac{{\rm E}(\sin(\Theta_i-\Theta_i^*)\sin(\Theta_{i^{\top}}-\Theta_{i^{\top}}^*)    )}{\sqrt{ {\rm E}(\sin^2(\Theta_i-\Theta_i^*)) {\rm E}(\sin^2(\Theta_{i^{\top}}-\Theta_{i^{\top}}^*)   )     }}  \in [-1,1]  
%,\label{eq:112}
%\end{equation}
%where $\Theta_i^*$ and $\Theta_{i^{\top}}^*$ are two circular variables  distributed, respectively, as $\Theta_i$ and $\Theta_{i^{\top}}$, and the  circular-linear dependence of \cite{mardia1976}, that is   
%\begin{align}
%\rho_{(\Theta_i,Y_{j})}^2& =\frac{Cor{(\cos\Theta_i,Y_{j} )}^2+{\rm Cor}{(\sin\Theta_i,Y_{j} )}^2    }{1-{\rm Cor}{(\cos\Theta_i,\sin\Theta_i )}}+\\
%& \frac{-2{\rm Cor}{(\cos\Theta_i,Y_{j} )}{\rm Cor}{(\sin\Theta_i,Y_{j} )}{\rm Cor}{(\cos\Theta_i,\sin\Theta_i )}    }{1-{\rm Cor}{(\cos\Theta_i,\sin\Theta_i )}}   \in [0,1]. \label{eq:111}
%\end{align}

 {
The MCMC algorithm is implemented using the same number of  iterations, thin and  burnin used in the previous section while ì $\{\boldsymbol{\mu},\boldsymbol{\Sigma}\}\sim\mathcal{NIW} \left( \mathbf{0}_{12}, 0.001, 15, \mathbf{I}_{12}  \right)$   and $\boldsymbol{\lambda} \sim \mathcal{N}_4\left( 0, 100 \mathbf{I}_4 \right)$. In Figure \ref{fig:ESTDlin2} and \ref{fig:Cor2} we depicted, respectively, the marginal  posterior $\mathcal{JPSN}$ densities
and  the dependence matrix. The latter  shows the MC estimates of the  posterior mean circular-circular correlation of \cite{Jammalamadaka1988}, 
the circular-linear dependence of  \cite{mardia1976}, i.e., 
\begin{align}
\rho_{(\Theta_i,Y_{j})}^2& =\frac{  {\rm Cor} {(\cos\Theta_i,Y_{j} )}^2+{\rm Cor}{(\sin\Theta_i,Y_{j} )}^2    }{1-{\rm Cor}{(\cos\Theta_i,\sin\Theta_i )}}+\\
& \frac{-2{\rm Cor}{(\cos\Theta_i,Y_{j} )}{\rm Cor}{(\sin\Theta_i,Y_{j} )}{\rm Cor}{(\cos\Theta_i,\sin\Theta_i )}    }{1-{\rm Cor}{(\cos\Theta_i,\sin\Theta_i )}}   \in [0,1], \label{eq:111}
\end{align}
 and     linear-linear correlation, evaluated with the Pearson's coefficient}. A circular-circular and linear-linear correlation is plotted only if the associated  CI does not {contain  zero} while
 {since the CI of  $\rho^2(\Theta_i,Y_j)$ has 0  probability  to contain the 0,}
 we plot the value of  $\rho^2(\Theta_i,Y_j)$ using a different rationality. More precisely, since  $\boldsymbol{W}_i\perp Y_j$ iff $\Theta_i \perp Y_j$, in Figure \ref{fig:Cor2} we plot the  posterior mean value of $\rho^2(\Theta_i,Y_j)$   only if at least one of the   CIs of   $\boldsymbol{\Sigma}_{wy}$ that measure  the correlation between $\boldsymbol{W}_i$ and $Y_j$ does not contain the zero.
From Figure  \ref{fig:ESTDlin2} we appreciate that the $\mathcal{JPSN}$ is able to fit satisfactorily the data and to find significant circular-linear  and linear-linear  correlations (Figure \ref{fig:Cor2}).

\subsection{Comparison with cylindrical distributions} \label{sec:compAbeLey}

 {
With this section we want to demonstrate that ignoring multivariate dependence leads to loss of  predictive ability.
Then we compare our proposal with the cylindrical distribution of Abe-Ley \citep{Abe2015} and a cylindrical version of the $\mathcal{JPSN}$, i.e.,  for both we assume  dependence between the  variables  belonging to the same animal and independence between zebras.
Since the $\mathcal{JPSN}$ is not available in closed form,  a comparison based on informational criteria, such as AIC or BIC, is not possible. We decide to make the  comparison  in terms of predictive ability} measured used   the CRPS, that is a proper scoring rule  % 
defined for both circular \citep{grimit2006} and linear  \citep{Gneiting2007} variables that measure the distance between cumulative distribution functions \citep{Matheson1976};   lower values are then preferable.
The Abe-Ley distribution is defined only for a positive linear variable and then, to make a fair comparison, we use the distribution that arises by taking the logarithm of its linear component:  { 
\begin{equation} \label{eq:ab}
f(\theta_i,y_j) =  \frac{\alpha^{AL} (\beta^{AL})^{\alpha^{AL}}}{2 \pi \cosh \kappa^{AL}}	\left( 1+\lambda^{AL} \sin \left(   \theta_i-\mu^{AL} \right)  \right) e^{y_j(\alpha^{AL} -1)}   e^{  - \left(  \beta^{AL} e^{y_j} \right)^{\alpha^{AL}}\left( 1-\tanh  \kappa^{AL}  \cos \left( \theta_i-\mu^{AL}  \right)  \right)  } e^{y_j} .
\end{equation}
$\alpha^{AL}\in \mathbb{R}^+$ and $\beta^{AL}\in \mathbb{R}^+$   are linear scale and shape parameters, $\mu^{AL} \in[0,2\pi)$ and $\lambda^{AL} \in [-1,1]$  endorse the role of circular
location and skewness parameters and $\kappa^{AL}\in \mathbb{R}^+$ plays  the role of circular concentration
and circular-linear dependence parameter. For the Abe-Ley parameters we use standard weak informative priors, i.e., an inverse gamma  with parameters (1,1) for  $\alpha^{AL}$, $\beta^{AL}$ and $\kappa^{AL}$ while uniform distributions on the respective domains are used for $\mu^{AL}$ and $\lambda^{AL}$. Under the cylindrical $\mathcal{JPSN}$, a $\mathcal{N}_1(0,100)$ is used for the skew parameters and a $\mathcal{NIW} \left( \mathbf{0}_{3}, 0.001, 6, \mathbf{I}_{3}  \right)$ for the others that are the marginal priors  deriving from the ones of the poly-cylindrical $\mathcal{JPSN}$.
}

We select 10\% of the circular and linear observations to be set aside and not used to estimate the posterior distributions. We predict their values based on the posterior samples and we measure how the  models perform in term of posterior estimates. 
 Then,  let $\mathcal{C}_i \subset \{1,2,\dots,T\}$ and $\mathcal{L}_j\subset \{1,\ldots,T\}$ be sets of indices, where $t \in \mathcal{C}_i$ if $\theta_{ti}$ is missing and 
$t \in \mathcal{L}_j$ if  $y_{tj}$ is missing, 
and let $\theta_{ti}^b$, $t \in \mathcal{C}_i$, and $y_{tj}^b$, $t \in \mathcal{L}_j$, be, respectively, the $b^{th}$ posterior sample of $\theta_{ti}$ and $y_{tj}$.  An MC approximation  of the CRPS for  circular variables based on $B$ posterior samples is computed as
\begin{equation}
CRPSc_{i} \approx \frac{1}{B}\sum_{b=1}^Bd(\theta_{ti},\theta_{ti}^b)-\frac{1}{2 B^2} \sum_{b=1}^B \sum_{b'=1}^B d(\theta_{ti}^b,\theta_{ti}^{b'}), \, t \in \mathcal{C}_i,
\end{equation}
where $d(\cdot,\cdot)$ is the angular distance, while  the CRPS for  linear variable is approximated by 
\begin{equation}
CRPSl_{j} \approx \frac{1}{B}\sum_{b=1}^B |y_{tj}-y_{tj}^b|-\frac{1}{2 B^2} \sum_{b=1}^B \sum_{b'=1}^B |y_{tj}^b-y_{tj}^{b'}|, \, t \in \mathcal{L}_j.
\end{equation}
We then compute the  overall mean CRPSs for the sets of circular  and linear variables
%, $CRPSc= \frac{1}{p}\sum_{i=1}^p CRPSc_{i}$, 
%, $CRPSl= \frac{1}{q}\sum_{j=1}^q CRPSl_{j}$, 
and we use these  indices to measure the goodness-of-fit.

 {
Circular and linear CRPSs  have  values  0.383
and  0.693 for the $\mathcal{JPSN}$, 0.385
and  0.762 for the cylindrical  $\mathcal{JPSN}$,
and    0.412
and 0.753 for the Abe-Ley density, showing that  the $\mathcal{JPSN}$ performs better and, moreover,  it is also able to give a measure of dependence between all the circular and linear components (Figure \ref{fig:Cor2}) that is not possible with cylindrical distributions.}

\section{Concluding remarks} \label{sec:conc}

In this work we introduced a  poly-cylindrical distribution. The proposal is highly flexible, it is closed under marginalization and  it allows  to have dependent components, bimodal marginal circular distributions and asymmetric linear ones. We showed how the MCMC algorithm, used to obtain posterior samples, can be easily implemented using only  Gibbs steps. 
The proposal suffers from an identification problem and we showed how to overcome it with a post-processing of posterior samples that can also be  used with the $\mathcal{PN}$ distribution.
With the aim to prove the validity of our sampling scheme,  the algorithm was  applied to  simulated examples.   
Then the proposed distribution was used to  model a real data taken from the movebank data repository. The predictive ability of our proposal was compared  with the ones of cylindrical distributions, showing that the $\mathcal{JPSN}$ performs better.

Future work will lead us to use the $\mathcal{JPSN}$ as emission distribution in an hidden Markov model and to 
incorporate covariates to model  mean and covariance  of the  circular-linear observations.

%
%One possible solution is to assume independence between $\tilde{\boldsymbol{\mu}}$ and $\tilde{\boldsymbol{\Sigma}} $  and to
%decompose  $\tilde{\boldsymbol{\Sigma}}$ into the correlation matrix ${\boldsymbol{\Omega}}$ and non-constrained variances. Then we can use  the prior proposed in \cite{Barnard00} for ${\boldsymbol{\Omega}}$ and, for example, inverse gamma priors for the variances.   Using the algorithm of \cite{Barnard00}, we can obtain posterior samples of ${\boldsymbol{\Omega}}$, but
%since the algorithm    draws  the elements of  ${\boldsymbol{\Omega}}$ one at a time, there is the danger that this approach   makes the MCMC algorithm slow to converge and  auto-correlated, while with  the approach we are going to introduce, 
%we  sample all together the elements of $\tilde{\boldsymbol{\Sigma}}$, making the algorithm more  efficient. 
%
%
%
%Other possibilities are to assume the   prior proposed by 
%\cite{Huang2013} or the one of \cite{omalley2008} for ${\boldsymbol{\Sigma}}$ and  a  $(2p+q)$-variate normal  for   ${\boldsymbol{\mu}}$.

\section*{Acknowledgement}
The author wishes to thank Antonello Maruotti, Giovanna Jona Lasinio  and Alessio Pollice for assistance and comments that greatly improved the manuscript.

This work is partially developed under the PRIN2015 supported-project ‘‘Environmental processes and human activities: capturing their interactions via statistical methods (EPHASTAT)’’ funded by MIUR (Italian Ministry of Education, University and Scientific Research).

%\end{acknowledgement}

\section*{References}
\bibliographystyle{model2-names}
\bibliography{all}
%\bibliography{all}

\appendix
\subsection*{Appendix}
\renewcommand{\thesubsection}{\Alph{subsection}}
\renewcommand{\theequation}{\Alph{subsection}.\arabic{equation}}

\subsection{The invariance property of the $\mathcal{PN}$} \label{sec:app1}

 {Here we prove that the univariate  marginal density of the circular variables  is invariant. 
Let $\Theta_i^*=\delta (\Theta_i+\xi)$, where $\delta \in \{-1,1\}$ and $\xi \in [0,2\pi)$,}
following Theorem 1 of \cite{mastrantonio2015f}, the density of $\Theta_i \sim \mathcal{PN}_1(\boldsymbol{\mu}_{w_i},\boldsymbol{\Sigma}_{w_i})$, i.e., $f_{\Theta_i}(\cdot)$, has the invariant property if $f_{\Theta_i^*}(\cdot)$, i.e., the density of $\Theta_i^*$, belongs to the same parametric family of $f_{\Theta_i}(\cdot)$.

The random variables $\Theta_i^*$ can be written as
\begin{equation} \label{eq:inv}
\Theta_i^* =\mbox{atan}^* \frac{\sin \Theta_i^*}{\cos \Theta_i^*} = \mbox{atan}^* \frac{\sin (\delta (\Theta_i+\xi))}{\cos (\delta (\Theta_i+\xi))} = \mbox{atan}^* \frac{ \delta \sin ( \Theta_i+\xi)}{\cos ( \Theta_i+\xi)} ,
\end{equation}
and using relations $\cos(\alpha+\beta) = \cos \alpha\cos\beta-\sin \alpha\sin \beta$ and $\sin(\alpha+\beta) = \sin \alpha\cos\beta+\cos \alpha\sin \beta$, equation \eqref{eq:inv} can be stated equivalently as
\begin{equation} 
\Theta_i^*  =    
\mbox{atan}^* \frac{ \delta (R_i \sin\Theta_i\cos\xi+ R_i\cos\Theta_i\sin \xi)   }{ R_i\cos \Theta_i\cos\xi-R_i\sin \Theta_i\sin \xi}  =  
\mbox{atan}^* \frac{ \delta (W_{i2}\cos\xi+W_{i1}\sin \xi)   }{ W_{i1}\cos\xi-W_{i2}\sin \xi}.
\end{equation}

To prove that $\Theta_i^*$ is $\mathcal{PN}$ distributed, let consider the random variable  $\mathbf{W}_i^*= \Delta\mathbf{T} \mathbf{W}_i$, where   $\Delta = \mbox{diag}((1,\delta)^{\top})$ and
\begin{equation}
\mathbf{T}= \left(
\begin{array}{cc}
\cos \xi &  -\sin \xi\\
\sin \xi & \cos \xi
\end{array}
\right).
\end{equation}
$\mathbf{W}_i^*$
 is normally distributed and  equation   \eqref{eq:tranpn} applied to $\mathbf{W}_i^*$ gives 
 \begin{equation} 
 \mbox{atan}^* \frac{W_{i2}^*   }{ W_{i1}^*}  = 
 \mbox{atan}^* \frac{ \delta (W_{i2}\cos\xi+W_{i1}\sin \xi)   }{ W_{i1}\cos\xi-W_{i2}\sin \xi} = \Theta_i^*.
 \end{equation}

Then  $\Theta_i^*$ follows a projected normal distribution; this proves the invariance of the $\mathcal{PN}$.

\subsection{MCMC implementation details} \label{sec:app2}

\paragraph{Sampling ${\boldsymbol{\mu}}$ and ${\boldsymbol{\Sigma}}$}

The full conditional of  $\{\boldsymbol{\mu},\boldsymbol{\Sigma} \}$ is  proportional to
\begin{equation} \label{eq:ww}
\prod_{t=1}^T\phi_{2p+q}((\mathbf{w}_t, \mathbf{y}_t)^{\top}-(\mathbf{0}_{2p},\text{diag}(\boldsymbol{\lambda})\mathbf{d}_t )^{\top} |{\boldsymbol{\mu}} , {\boldsymbol{\Sigma}})  f(\boldsymbol{\mu},\boldsymbol{\Sigma}|\boldsymbol{\lambda}).
\end{equation}
Equation \eqref{eq:ww}  is equivalent to the full conditional of the mean and covariance matrix  in a model with  i.i.d. normally distributed observations.
If we assume $f(\boldsymbol{\mu},\boldsymbol{\Sigma}|\boldsymbol{\lambda}) \equiv f(\boldsymbol{\mu},\boldsymbol{\Sigma})  $, with 
\begin{equation}
f(\boldsymbol{\mu},\boldsymbol{\Sigma})   \propto \left| \boldsymbol{\Sigma}  \right|^{-  (\nu_0+2p+q)/2-1     } \exp \left(  { - \frac{ {\rm tr}\left(\boldsymbol{\Psi}_0  \boldsymbol{\Sigma}^{-1}\right)+\kappa_0\left(  \boldsymbol{\mu}-\boldsymbol{\mu}_0  \right)^{\top}   \boldsymbol{\Sigma}^{-1}    \left(  \boldsymbol{\mu}-\boldsymbol{\mu}_0  \right)  }{2}  }\right)
\end{equation}
i.e.,  $f(\boldsymbol{\mu},\boldsymbol{\Sigma}) $ is the density of a $\mathcal{NIW}(\boldsymbol{\mu}_0, \kappa_0,\nu_0,\boldsymbol{\Psi}_0)$,  where $\kappa_0>0$ and $\nu_0>2p+q-1$ are real numbers,
 $\boldsymbol{\mu}_0 \in \mathbb{R}^{2p+q}$ and $\boldsymbol{\Psi}_0$ is a $(2p+q)\times(2p+q)$ nnd matrix, and we let $\boldsymbol{\eta}_t= (\mathbf{w}_t, \mathbf{y}_t)^{\top}-(\mathbf{0}_{2p},\text{diag}(\boldsymbol{\lambda})\mathbf{d}_t )^{\top}$
and
 \begin{equation}
\bar{\boldsymbol{\eta}} =	\frac{1}{T} \sum_{t=1}^T\boldsymbol{\eta}_t,
 \end{equation}
 the full conditional is $\mathcal{NIW}(\boldsymbol{\mu}_{\text{post}}, \kappa_{\text{post}},\nu_{\text{post}},\boldsymbol{\Psi}_{\text{post}})$ with 
 \begin{align}
\boldsymbol{\mu}_{\text{post}} & =  \frac{\kappa_0 \boldsymbol{\mu}_0+T \bar{\boldsymbol{\eta}}  }{\kappa_0+T},\\
\kappa_{\text{post}} & =  \kappa_0+T,\\
\nu_{\text{post}} & =  \nu_0+T,\\
\boldsymbol{\Psi}_{\text{post}} & =  \boldsymbol{\Psi}_0+\sum_{t=1}^T\left( {\boldsymbol{\eta}_t} -\bar{\boldsymbol{\eta}}    \right)\left( {\boldsymbol{\eta}_t} -\bar{\boldsymbol{\eta}}    \right)^{\top}+\frac{\kappa_0T}{\kappa_0+T}\left( \bar{\boldsymbol{\eta}}-\boldsymbol{\mu}_0   \right)\left( \bar{\boldsymbol{\eta}}-\boldsymbol{\mu}_0   \right)^{\top}.
 \end{align}

\paragraph{Sampling $\boldsymbol{\lambda}$}
The full conditional of  $\boldsymbol{\lambda}$ is  proportional to
\begin{equation} \label{eq:fulllambda}
\prod_{t=1}^T \phi_{q}(\mathbf{y}_t |\boldsymbol{\mu}_{y_t|w_t}+ \text{diag}(\mathbf{d}_t) \boldsymbol{\lambda}   , \boldsymbol{\Sigma}_{y|w} ) g_2(\boldsymbol{\lambda}),
\end{equation}
where $\boldsymbol{\mu}_{y_t|w_t} =\boldsymbol{\mu}_{y}+\boldsymbol{\Sigma}_{wy}^{\top}\boldsymbol{\Sigma}_{w}^{-1}\left( \mathbf{w}_t - \boldsymbol{\mu}_{w} \right) $ and   $\boldsymbol{\Sigma}_{y|w} = \boldsymbol{\Sigma}_{y}- \boldsymbol{\Sigma}_{wy}^{\top}\boldsymbol{\Sigma}_{w}^{-1}\boldsymbol{\Sigma}_{wy}$.    In \eqref{eq:fulllambda} we can see $\boldsymbol{\lambda}$ as a  vector of regression coefficients, where the matrix of covariates is  $\text{diag}(\mathbf{d}_t)$. Then, standard results tell us that a normal $g_2(\boldsymbol{\lambda})$ induces a  normal full conditional. More precisely, let $\boldsymbol{\lambda}  \sim \mathcal{N}_{q}( \boldsymbol{\gamma}_0,\boldsymbol{\Omega}_0 )$, then the full conditional is $ \mathcal{N}_{q}( \boldsymbol{\gamma}_{\text{post}},\boldsymbol{\Omega}_{\text{post}} )$ with
\begin{align}
\boldsymbol{\Omega}_{\text{post}} & = \left( \sum_{t=1}^T\text{diag}(\mathbf{d}_t)\boldsymbol{\Sigma}_{y|w}^{-1}  \text{diag}(\mathbf{d}_t)+   \boldsymbol{\Omega}_0^{-1}   \right)^{-1},\\
\boldsymbol{\gamma}_{\text{post}} & = \boldsymbol{\Omega}_{\text{post}}\left(  \sum_{t=1}^T\text{diag}(\mathbf{d}_t)\boldsymbol{\Sigma}_{y|w}^{-1}  \left(   \mathbf{y}_t -\boldsymbol{\mu}_{y_t|w_t}\right) +\boldsymbol{\Omega}_0^{-1}\boldsymbol{\gamma}_0   \right).
\end{align}

\paragraph{Sampling $\mathbf{D}_t$}
The full conditional of the latent vector  $\mathbf{D}_t$ is  proportional to
\begin{equation} \label{eq:fullDt}
\phi_{q}(\mathbf{y}_t |\boldsymbol{\mu}_{y_t|w_t}+ \mbox{diag}(\boldsymbol{\lambda})\mathbf{d}_t    , \boldsymbol{\Sigma}_{y|w} )  \phi_q(\mathbf{d}_t| \mathbf{0}_q,\mathbf{I}_q ) .
\end{equation}
$\mathbf{d}_t$ can be seen as a vector of (positive) regressors with $\mbox{diag}(\boldsymbol{\lambda})$ as matrix of  covariates  and  $\phi_q(\mathbf{d}_t| \mathbf{0}_q,\mathbf{I}_q )$   as  prior. The full conditional is  then
$N_{q}\left(\mathbf{M}_{d_t}, \mathbf{V}_q  \right)I_{\mathbf{0_q, \boldsymbol{\infty}}}$, where $N_{q}\left(\cdot, \cdot  \right)I_{\mathbf{0_q, \boldsymbol{\infty}}}$ is a  $q-$dimensional truncated normal distribution with  components having support $\mathbb{R}^+$, 
\begin{align}
\mathbf{V}_{d} &=\left(\boldsymbol{\Lambda}^{\top} \boldsymbol{\Sigma}_{y|w}^{-1}\boldsymbol{\Lambda}+ \mathbf{I}_{q}  \right)^{-1}
\end{align}
and
\begin{align}
\mathbf{M}_{d_t}& = \mathbf{V}_d\boldsymbol{\Lambda}^{\top} \boldsymbol{\Sigma}_{y|w}^{-1}   \left(\mathbf{y}_{t}- \boldsymbol{\mu}_{y_t|w_t}\right).
\end{align}

\paragraph{Sampling $R_{ti}$}

Let $\mathbf{u}_{ti} = \left(  \cos \theta_{ti},\sin \theta_{ti} \right)^{\top}$, $A_{ti} = \mathbf{u}_{ti}^{\top} \boldsymbol{\Sigma}_{w_{ti}|w_{t-i},y}^{-1} \mathbf{u}_{ti}$ and $B_{ti} =  \mathbf{u}_{ti}^{\top} \boldsymbol{\Sigma}_{w_{ti}|w_{t-i},y}^{-1}\boldsymbol{\mu}_{w_{ti}|w_{t-i},y_t}$, where   $\boldsymbol{\mu}_{w_{ti}|w_{t-i},y_t}$ and $\boldsymbol{\Sigma}_{w_{ti}|w_{t-i},y}$ are the conditional mean and covariance matrix of $\mathbf{w}_{ti}$ assuming $(\mathbf{w}_t, \mathbf{y}_t )^{\top} \sim \mathcal{N}_{2p+q}\left( \boldsymbol{\mu}+(\mathbf{0}_{2p},\text{diag}(\boldsymbol{\lambda})\mathbf{d}_t )^{\top} , {\boldsymbol{\Sigma}}   \right)$.
The full conditional of $R_{ti}$ is then proportional to
\begin{equation} 
r_{ti} \exp \left(   -\frac{1}{2}  A_{ti} \left( r_{ti} - \frac{B_{ti}}{A_{ti}}  \right)^2 \right). \label{eq:r}
\end{equation}

Equation \eqref{eq:r} is the same full conditional  of the latent variable of  the   spherical $\mathcal{PN}$ of \cite{Stumpfhause2016} and then we can  use their  \emph{slice sampling} strategy to sample from it.

In details, if
\begin{align}
v_{ti} &\sim \mathcal{U}\left( 0,   \exp \left(   -\frac{1}{2}  A_{ti} \left( r_{ti} - \frac{B_{ti}}{A_{ti}}  \right)^2 \right) \right),\\
v_{ti}^*& \sim \mathcal{U}(0,1),
\end{align}
then 
\begin{equation}
r_{ti}  =    \sqrt{       \left(      \varrho_{2ti}^2- \varrho_{1ti}^2  \right)   v_{ti}^*        + \varrho_{1ti}^2      }  ,  
\end{equation}
with 
\begin{align}
 \varrho_{1ti} =&   \frac{B_{ti}}{A_{ti}}+ {\rm max} \left\{      -\frac{B_{ti}}{A_{ti}},- \sqrt{ \frac{-2 \ln v_{ti} }{A_{ti}}  }    \right\},\\
  \varrho_{2ti} =&   \frac{B_{ti}}{A_{ti}}+\sqrt{ \frac{-2 \ln v_{ti} }{A_{ti}}  },
\end{align}
is distributed accordingly to the full conditional  \eqref{eq:r}.

\end{document}